\pdfoutput=1
\documentclass[12pt,a4paper]{article}

\usepackage{ifthen} 
\newboolean{pdflatex}
\setboolean{pdflatex}{true} 

\newboolean{articletitles}
\setboolean{articletitles}{true} 

\newboolean{uprightparticles}
\setboolean{uprightparticles}{false} 

\def\paperauthors{Federico Betti} 
\def\paperasciititle{A Review of CP Violation Measurements in Charm at LHCb} 
\def\papertitle{A Review of $C\!P$ Violation Measurements in Charm at LHCb} 

\usepackage[top=1in, bottom=1.25in, left=1in, right=1in]{geometry}

\usepackage{microtype}
\usepackage{lineno}  
\usepackage{xspace} 
\usepackage{caption} 

\usepackage{graphicx}  
\usepackage{color}
\usepackage{colortbl}
\graphicspath{{./figs/}} 

\usepackage{amsmath} 
\usepackage{amssymb}
\usepackage{amsfonts}
\usepackage{upgreek} 

\newcommand*\patchAmsMathEnvironmentForLineno[1]{%
\expandafter\let\csname old#1\expandafter\endcsname\csname #1\endcsname
\expandafter\let\csname oldend#1\expandafter\endcsname\csname
end#1\endcsname
 \renewenvironment{#1}%
   {\linenomath\csname old#1\endcsname}%
   {\csname oldend#1\endcsname\endlinenomath}%
}
\newcommand*\patchBothAmsMathEnvironmentsForLineno[1]{%
  \patchAmsMathEnvironmentForLineno{#1}%
  \patchAmsMathEnvironmentForLineno{#1*}%
}
\AtBeginDocument{%
\patchBothAmsMathEnvironmentsForLineno{equation}%
\patchBothAmsMathEnvironmentsForLineno{align}%
\patchBothAmsMathEnvironmentsForLineno{flalign}%
\patchBothAmsMathEnvironmentsForLineno{alignat}%
\patchBothAmsMathEnvironmentsForLineno{gather}%
\patchBothAmsMathEnvironmentsForLineno{multline}%
\patchBothAmsMathEnvironmentsForLineno{eqnarray}%
}

\usepackage{hyperxmp}

\usepackage[pdftex,
            pdfauthor={\paperauthors},
            pdftitle={\paperasciititle},
            ]{hyperref}

\usepackage[colorinlistoftodos,textsize=scriptsize]{todonotes}

\usepackage[bottom,flushmargin,hang,multiple]{footmisc}

\usepackage[all]{hypcap} 

\usepackage{cite} 
\usepackage{mciteplus}

\begin{document}

\renewcommand{\thefootnote}{\fnsymbol{footnote}}
\setcounter{footnote}{1}

\begin{titlepage}
\pagenumbering{roman}

\vspace*{-1.5cm}
\vspace*{1.5cm}
\noindent

\vspace*{4.0cm}

{\normalfont\bfseries\boldmath\huge
\begin{center}
  \papertitle 
\end{center}
}

\vspace*{2.0cm}

\begin{center}
Federico Betti\footnote{federico.betti@cern.ch} \\
\textit{CERN (European Organization for Nuclear Research) \\
Espl. des Particules 1, 1211 Meyrin, Switzerland}
\end{center}

\vspace{\fill}

\begin{abstract}
  \noindent
  The LHCb experiment has been able to collect the largest sample ever produced of charm-hadron decays, performing a number of measurements of observables related to $C\!P$ violation in the charm sector. In this document, the most recent results from LHCb on the search of direct $C\!P$ violation in $D^0 \to K_s^0 K_s^0$, $D_{(s)}^+ \to h^+ \pi^0$ and $D_{(s)}^+ \to h^+ \eta$ decays are summarised, in addition to the most precise measurement of time-dependent $C\!P$ asymmetry in $D^0 \to h^+ h^-$ decays and the first observation of mass difference between neutral charm-meson eigenstates.
\end{abstract}

\vspace*{2.0cm}

\begin{center}
  Published in Symmetry \textbf{13} (2021) 8, 1482
\end{center}

\vspace{\fill}

\vspace*{2mm}

\end{titlepage}

\newpage
\setcounter{page}{2}
\mbox{~}

\renewcommand{\thefootnote}{\arabic{footnote}}
\setcounter{footnote}{0}

\cleardoublepage

\pagestyle{plain} 
\setcounter{page}{1}
\pagenumbering{arabic}

\section{Introduction}

For a generic hadron $M$ and its $C\!P$ conjugate $\overline{M}$, the~following decay amplitudes in the final state $f$ and its $C\!P$ conjugate $\overline{f}$ are defined
\begin{equation}
    A_f = \langle f | \mathcal{H} | M \rangle,\qquad \overline{A}_{\overline{f}} = \langle \overline{f} | \mathcal{H} | \overline{M} \rangle,
\label{amplitudes}
\end{equation}
where $\mathcal{H}$ is the effective decay Hamiltonian.
The $C\!P$ asymmetry of the decay $M \to f$ is therefore defined as
\begin{equation}
    \mathcal{A}_{C\!P}(M \to f) \equiv \frac{\Gamma(M \to f) - \Gamma(\overline{M} \to \overline{f})}{\Gamma(M \to f) + \Gamma(\overline{M} \to \overline{f})} = \frac{1 - | \overline{A}_{\overline{f}}/A_f |^2}{1 + | \overline{A}_{\overline{f}}/A_f |^2}.
\label{ACP}
\end{equation}

Direct $C\!P$ violation, or~$C\!P$ violation in the decay, takes place when $ |\overline{A}_{\overline{f}} |^2 \neq |A_f |^2$.
For neutral mesons such as $D^0$, mixing must be taken into account.
The eigenstates $|D_H\rangle$ and $|D_L\rangle$ of the effective Hamiltonian, which have defined masses ($m_H$ and $m_L$) and decay widths ($\Gamma_H$ and $\Gamma_L$), can be written as a superposition of flavour eigenstates
\begin{align}
    |D_H\rangle &= p|D^0\rangle - q|\overline{D}^0 \rangle, \\
    |D_L\rangle &= p|D^0\rangle + q|\overline{D}^0 \rangle, 
\label{mass_eig}
\end{align}
where $p$ and $q$ are complex with $|p|^2 + |q|^2 = 1$.
The mixing is described by the parameters
\begin{align}
    x &\equiv \frac{m_H - m_L}{\Gamma}, \\
    y &\equiv \frac{\Gamma_H - \Gamma_L}{2\Gamma},
\label{mix_pars}
\end{align}
with $\Gamma = (\Gamma_H + \Gamma_L) / 2$.
If $|q/p| \neq 1$, $C\!P$ violation in the mixing occurs, and~the probability of the $D^0$ meson to oscillate after a time $t$ into $\overline{D}^0$ is different from that of the $C\!P$-conjugate process.
If $f = \overline{f}$ and the phase
\begin{equation}
    \phi_f \equiv \operatorname{arg}\left( \frac{q \overline{A}_f}{p A_f} \right)
\label{phase_interf}
\end{equation}
is different from 0, $C\!P$ violation manifests in the interference between mixing and decay.
The time-dependent $C\!P$ asymmetry between the probability of an initially pure $D^0$ meson and an initially pure $\overline{D}^0$ meson to decay into the final state $f$ after a time $t$
\begin{equation}
    \mathcal{A}_{C\!P}(D^0(t) \to f) \equiv \frac{\Gamma(D^0(t) \to f) - \Gamma(\overline{D}^0(t) \to f)}{\Gamma(D^0(t) \to f) + \Gamma(\overline{D}^0(t) \to f)}
\label{ACP_td}
\end{equation}
is different from 0 if $C\!P$ violation occurs either in the mixing or in the interference between mixing and decay.
In the $D^0$ meson system, the~time-dependent $C\!P$ asymmetry can be written as
\begin{equation}
    \mathcal{A}_{C\!P}(D^0(t) \to f) \approx a^d_f + \frac{4 |A_f|^2 |\overline{A}_f|^2}{(|A_f|^2+|\overline{A}_f|^2)^2} \Delta Y_f \frac{t}{\tau_{D^0}},
\label{ACP_td_approx}
\end{equation}
where $a^d_f$ is equal to the direct $C\!P$ asymmetry, $\tau_{D^0}$ is the $D^0$ lifetime and $\Delta Y_f$, the~observable describing the amount of time-dependent $C\!P$ violation, is defined according to the ``theoretical'' parametrisation introduced in Refs.~\cite{Grossman:2009mn,Kagan:2009gb}.
The coefficient in front of $\Delta Y_f$ differs from unity by approximately $\mathcal{O}(10^{-6})$, resulting in $\Delta Y_f$ to be equal to the slope of $\mathcal{A}_{C\!P}(D^0(t) \to f)$.

The combination of Cabibbo--Kobayashi--Maskawa matrix elements responsible for $C\!P$ violation in charm decays is $\operatorname{Im}(V_{cb}V_{ub}^* / V_{cs}V_{us}^*) \simeq -6 \times 10^{-4}$, resulting in asymmetries typically of the order of $10^{-4}$--10$^{-3}$ in the Standard Model (SM).
Due to the presence of low-energy strong-interaction effects, theoretical predictions are difficult to compute reliably~\cite{Golden:1989qx,Buccella:1994nf,Bianco:2003vb,Artuso:2008vf,Brod:2011re,Cheng:2012wr,Cheng:2012xb,Li:2012cfa,Franco:2012ck,Pirtskhalava:2011va,Feldmann:2012js,Brod:2012ud,Hiller:2012xm,Grossman:2012ry,Bhattacharya:2012ah,Muller:2015rna,Khodjamirian:2017zdu,Buccella:2019kpn}.
Some calculations are performed using dynamical methods of QCD, such as Light-Cone Sum Rules~\cite{Cheng:2012wr,Li:2012cfa,Khodjamirian:2017zdu}.
Other studies rely on perturbative parameterisations of the branching fractions and the $C\!P$ asymmetries as a function of the topological amplitudes contributing to the dominant and subleading amplitudes in $SU(3)_F$ or $U$-spin breaking, fitting them to the measured values~\cite{Golden:1989qx,Buccella:1994nf,Franco:2012ck,Pirtskhalava:2011va,Feldmann:2012js,Brod:2012ud,Hiller:2012xm,Grossman:2012ry,Muller:2015rna,Buccella:2019kpn}.

Almost all the measurements of $C\!P$ violation in charm hadrons have been performed at the $B$-factories Belle and BaBar and at the hadron-collider experiments LHCb and CDF.
The production cross-section of charm in proton--proton ($pp$) (or proton--antiproton) collisions is much larger than that at the $B$-factories, for~example $\sigma(pp \to c \overline{c}X) = (2369 \pm 12)~\upmu\mathrm{b}$ at $\sqrt{s} = 13~\mathrm{TeV}$ in the LHCb acceptance~\cite{LHCb:2015swx}, whereas $\sigma(e^+ e^- \to \gamma^* \to c \overline{c}) = 1.3~\mathrm{nb}$ at $\sqrt{s} = 10.58~\mathrm{GeV}$.
Despite the large hadronic background, not present in the low-multiplicity environment of the $e^+ e^-$ machines, LHCb was therefore able to collect the largest sample ever produced of charm-hadron decays, performing a number of measurements of observables related to $C\!P$ violation in the charm sector, such as: searches for direct $C\!P$ asymmetry in $D_{(s)}^+$ decays~\cite{Aaij:2014qec,Aaij:2017eux,Aaij:2019vnt}; time-dependent $C\!P$ violation in singly Cabibbo suppressed (SCS) $D^0$ decays~\cite{Aaij:2015yda,Aaij:2017idz,Aaij:2018qiw,Aaij:2019yas}; direct $C\!P$ asymmetries in SCS $D^0$ decays~\cite{Aaij:2014gsa,Aaij:2016cfh,LHCb:2016nxk}; $C\!P$ violation in mixing using $D^0 \to K^+ \pi^-$ decays~\cite{Aaij:2016roz,Aaij:2017urz}; $C\!P$ violation in multi-body $D^0$ decays~\cite{Aaij:2013swa,Aaij:2013jxa,Aaij:2014qwa,Aaij:2014afa,Aaij:2018fpa,Aaij:2018nis}; $C\!P$ violation in $D^0$ decays into final states with $K_s^0$~\cite{Aaij:2018jud,Aaij:2019jot}; and $C\!P$ violation in baryon decays~\cite{LHCb:2017hwf,Aaij:2020wil}.
The measurement of $\Delta \mathcal{A}_{C\!P} \equiv \mathcal{A}_{C\!P}(D^0 \to K^+ K^-) - \mathcal{A}_{C\!P}(D^0 \to \pi^+ \pi^-)$ with the full data sample collected by the LHCb detector led to the first observation of $C\!P$ violation in the decay of charm hadrons~\cite{Aaij:2019kcg}.
The result, that challenges perturbative and sum-rule estimates of $\Delta \mathcal{A}_{C\!P}$~\cite{Grossman:2006jg,Khodjamirian:2017zdu}, prompted a renewed interest of the theory community in the field, sparking a discussion whether the signal is consistent with the Standard Model or if it is a sign of new physics~\cite{Chala:2019fdb,Grossman:2019xcj,Li:2019hho,Soni:2019xko,Cheng:2019ggx,Dery:2019ysp,Wang:2020gmn,Bause:2020obd,Dery:2021mll,Cheng:2021yrn,Kagan:2020vri}.
While according to some studies the discrepancy is due to an enhancement of rescattering beyond expectation~\cite{Pirtskhalava:2011va,Feldmann:2012js,Brod:2012ud,Grossman:2012ry,Muller:2015rna,Grossman:2019xcj,Soni:2019xko}, for~other authors the reason of the discrepancy must be found in new interactions beyond the SM~\cite{Chala:2019fdb}.
Possible contributions of new physics to $\Delta \mathcal{A}_{C\!P}$, such as models involving $Z'$, have been studied~\cite{Chala:2019fdb,Dery:2019ysp,Bause:2020obd}.

Additional investigations are therefore crucial to clarify the picture and solve open theoretical puzzles.
More precise measurements of branching fractions and $C\!P$ asymmetries will allow the sum rules relating the $C\!P$ asymmetries of different decay channels to be tested, and~a better precision on the predictions obtained with the topological amplitudes to be reached~\cite{Grossman:2012ry,Muller:2015rna,Li:2019hho,Cheng:2019ggx}.

In the following Section, the~most recent results obtained by the LHCb collaboration in the field of $C\!P$ violation in charm will be summarised, namely the measurements of: $C\!P$ asymmetry in $D^0 \to K_s^0 K_s^0$ decays; $C\!P$ asymmetry in $D_{(s)}^+ \to h^+ \pi^0$ and $D_{(s)}^+ \to h^+ \eta$ decays, where $h^+$ denotes a $\pi^+$ or $K^+$ meson; $\Delta Y$ in $D^0 \to K^+ K^-$ and $D^0 \to \pi^+ \pi^-$ decays; and mixing parameters by using $D^0 \to K_s^0 \pi^+ \pi^-$ decays.
In Section~\ref{sec_conclusions}, the~conclusions and the prospects for future measurements of $C\!P$ violation in charm at LHCb are~given.

\section{Recent results from LHCb}

The LHCb detector~\cite{Alves:2008zz,LHCb:2014set} is a single-arm forward spectrometer that covers the pseudorapidity range between 2 and 5, whose main goal is the study of particles containing $b$ or $c$ quarks.
The tracking system consists of a silicon-strip vertex detector (VELO), a~large-area silicon-strip detector, a~dipole magnet and three stations of silicon-strip detectors and straw drift tubes.
The momentum of charged particles is measured with a relative uncertainty varying from 0.5\% to 1.0\%.
The impact parameter (IP) of a track, defined as its minimum distance to a primary $pp$ collision vertex (PV), is measured with a resolution of $(15 \pm 29/p_\mathrm{T})~\upmu\mathrm{m}$, where $p_\mathrm{T}$ is the momentum transverse to the beam, expressed in $\mathrm{GeV}/c$.
Particle identification of charged hadrons is performed using information from two ring-imaging Cherenkov detectors.
Electrons, photons and hadrons are identified by a calorimeter system made of scintillating-pad and preshower detectors, an~electromagnetic and a hadronic calorimeter.
Muons are identified by a system composed of alternating layers of iron and multiwire proportional chambers.
The online event selection is performed by a trigger, which consists of a hardware stage followed by a software stage.
The hardware stage is based on information from the calorimeter and muon systems, whereas the software stage applies a full event~reconstruction.

In the measurements of $C\!P$ asymmetries in the decays of neutral $D^0$ mesons to $C\!P$ eigenstates, the~$D^0$ flavour, when produced, is determined by the charge of the accompanying ``tagging'' pion in the strong $D^{*+} \to D^0 \pi^+$ (Inclusion
 of the charge-conjugate process is implied throughout this document unless explicitly specified.) decay promptly produced in the $pp$ collisions, or~by the charge of the muon in the semileptonic $b$-hadron secondary decays.
Due to the higher production cross section of charm with respect to that of beauty, the~data samples selected with prompt tag usually convey a signal yield which is a factor $\sim$5 larger than that with semileptonic~tag.

The ``raw'' asymmetry obtained by counting the positively- and negatively-tagged signal candidates is not equal to the physical asymmetry $\mathcal{A}_{C\!P}$, but~it includes also experimental effects due to production asymmetry $\mathcal{A}_\mathrm{P}$ and detection asymmetry $\mathcal{A}_\mathrm{D}$.
The production asymmetry is due to different production rates of the charm meson and its anti-meson, whereas different interaction cross-sections of particle and antiparticles with the detector, which is not perfectly symmetric, result in the detection asymmetry.
These nuisance asymmetries, averaged over phase space for selected events, are usually $\mathcal{O}(10^{-2})$ or less~\cite{LHCb:2017bdt,LHCb:2013sxc,LHCb:2012fb,LHCb:2012swq}, allowing the measured raw asymmetry $\mathcal{A}_\mathrm{raw}$ to be written as a sum of the physical and the experimental asymmetries.
For example, the~raw asymmetry of promptly-tagged $D^0$ mesons that decay in a $C\!P$ eigenstate $f$ can be approximated as
\begin{equation}
    \mathcal{A}_\mathrm{raw}(D^0 \to f) \approx \mathcal{A}_{C\!P}(D^0 \to f) + \mathcal{A}_\mathrm{P}(D^{*+}) + \mathcal{A}_\mathrm{D}(\pi^+),
\label{Araw}
\end{equation}
where $\mathcal{A}_\mathrm{P}(D^{*+})$ is the production asymmetry of the $D^{*+}$ meson and $\mathcal{A}_\mathrm{D}(\pi^+)$ is the detection asymmetry of the tagging pion.
The approximation is valid up to corrections of $\mathcal{O}(10^{-6})$.
Calibration samples which share the same production and detection asymmetries of the signal sample are typically used to cancel the nuisance asymmetries and obtain the $C\!P$ asymmetry.

\subsection{Measurement of $C\!P$ asymmetry in $D^0 \to K_s^0 K_s^0$ decays}
\label{sec_KSKS}

In the $D^0 \to K_s^0 K_s^0$ decay, the~only contributing amplitudes proceed via tree-level exchange and loop-suppressed diagrams which are of similar size and vanish in the flavour-SU(3) limit.
Their interference could result in a $C\!P$ asymmetry up to the percent level in the SM~\cite{Nierste:2015zra}.
The CLEO, Belle and LHCb (with data collected during the LHC Run~1) collaborations have measured in the past the $C\!P$ asymmetry $\mathcal{A}_{C\!P}(D^0 \to K_s^0 K_s^0)$~\cite{Bonvicini:2000qm,Dash:2017heu,Aaij:2015fua}, with~the best precision reached by Belle, at~the level of 1.5\%.
The update performed by LHCb with data collected during the LHC Run~2, corresponding to an integrated luminosity of $6~\mathrm{fb}^{-1}$, is reported~here.

From an experimental point of view, the~relatively large lifetime of $K_s^0$ meson is challenging, because~the majority of $K_s^0$ mesons decay after the LHCb Vertex Locator and their decay products can be reconstructed only in the tracking stations further downstream along the beam axis.
The  first  software  trigger stage is designed to accept the $K_s^0$ mesons whose pions are reconstructed also in the Vertex Locator (referred to as ``long'' and indicated by ``L'' in the following), resulting in a worse trigger efficiency for $K_s^0$ whose pions are reconstructed only in the downstream part of the tracking system (referred to as ``D'').
Depending on how the decay products of the two $K_s^0$ mesons are reconstructed, the~data sample is split in three categories that are analysed separately: two long (LL), one long and one downstream (LD) and two downstream (DD) $K_s^0$ mesons.
Once the two $K_s^0$ mesons are combined in a $D^0$ candidate, its flavour is determined by looking at the charge of the accompanying pion in the reconstructed strong decay $D^{*+} \to D^0 \pi^+$.
The data sample is further split in two categories, according to whether the $D^{*+}$ candidate is compatible or incompatible with having originated from the PV.
The PV compatibility classification is not performed for the DD sample, because~of its low statistics and signal purity.
A multivariate classifier sensitive to the signal-to-background ratio is used to select high signal purity $D^{*+}$ candidates.
Finally, data collected in 2015 and 2016 are analysed separately from those collected in 2017--2018, because~of differences in trigger requirements between these two~periods.

\begin{figure}[h]
\centering
\includegraphics[width=6.5 cm]{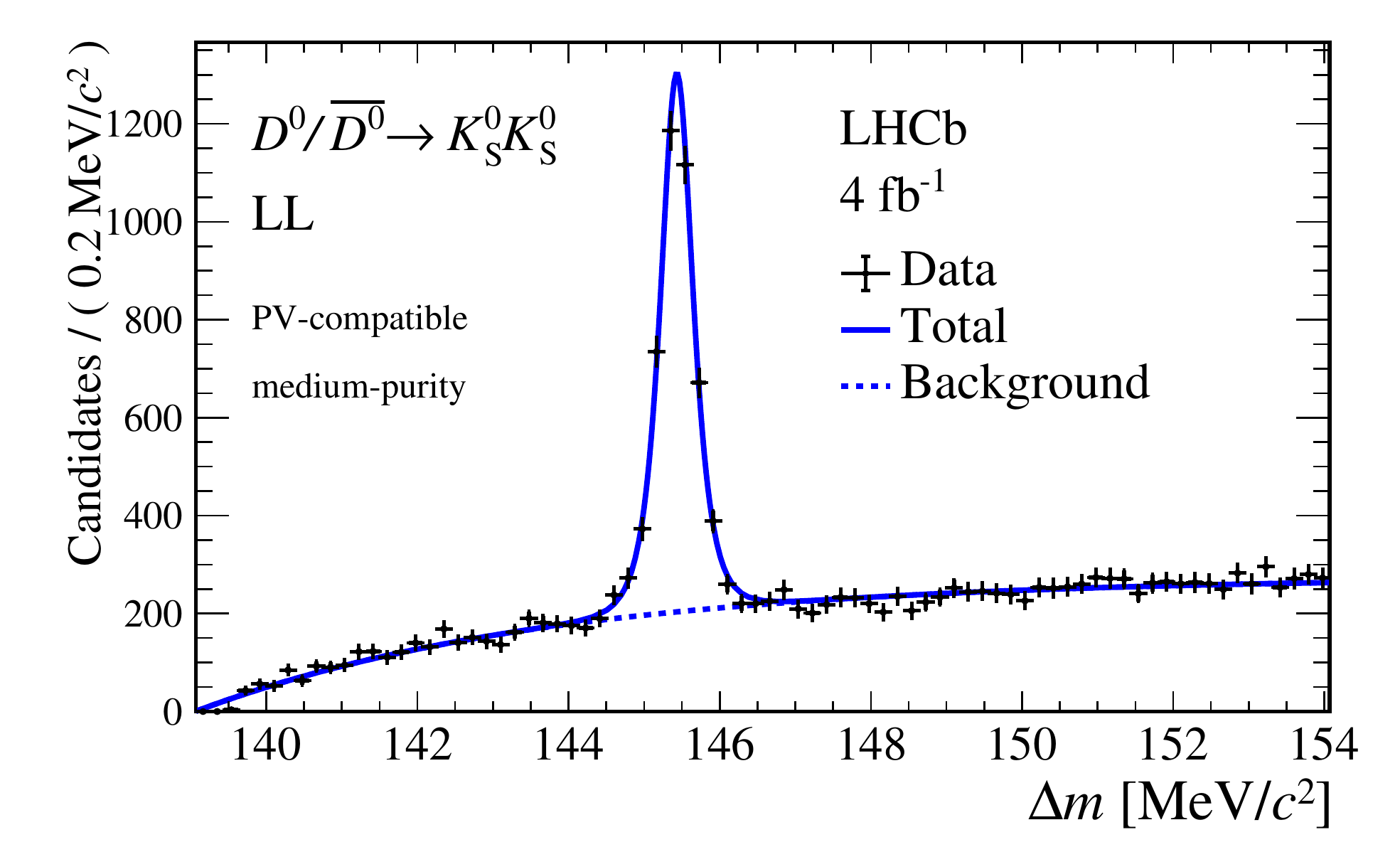}
\includegraphics[width=6.5 cm]{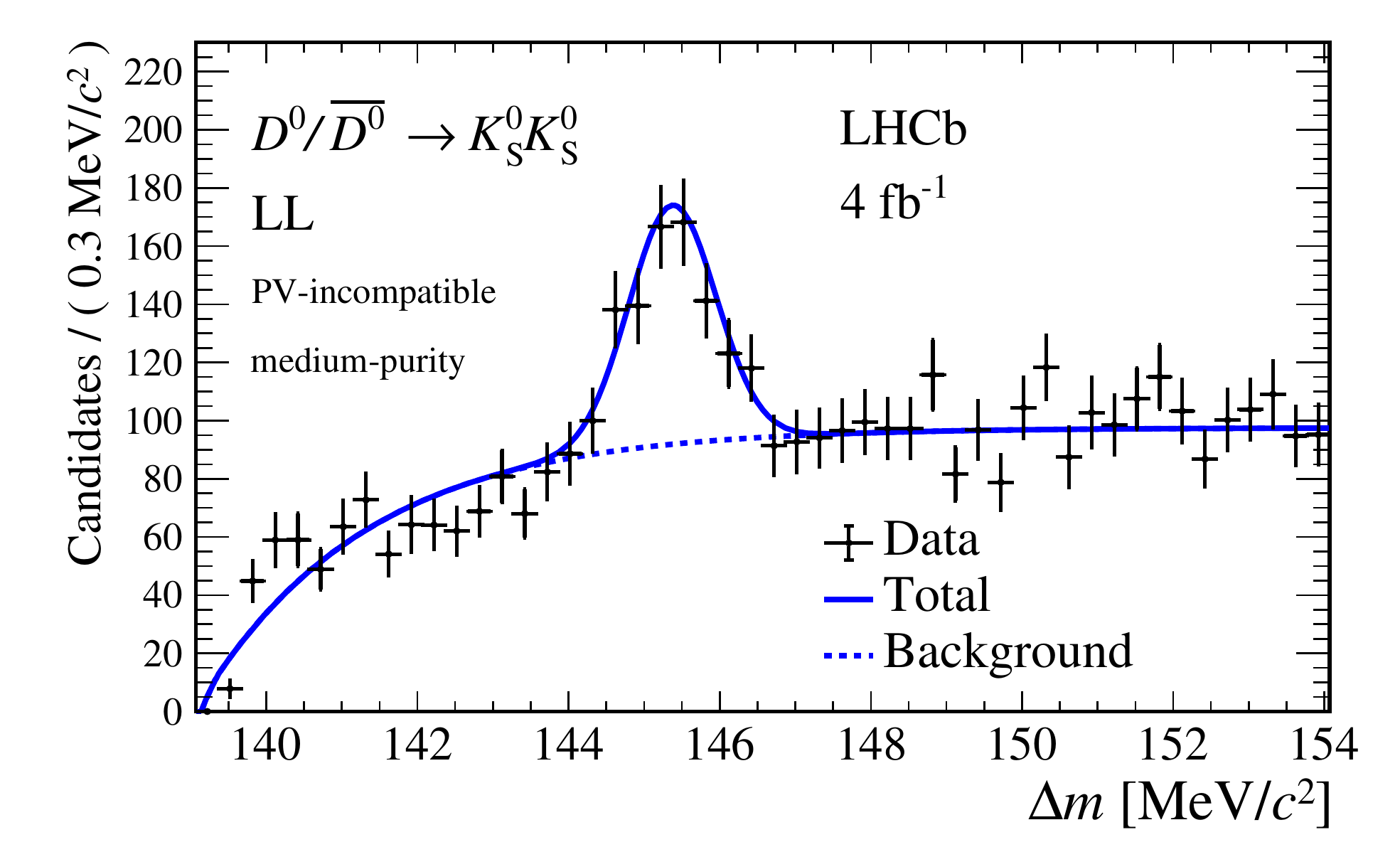}
\includegraphics[width=6.5 cm]{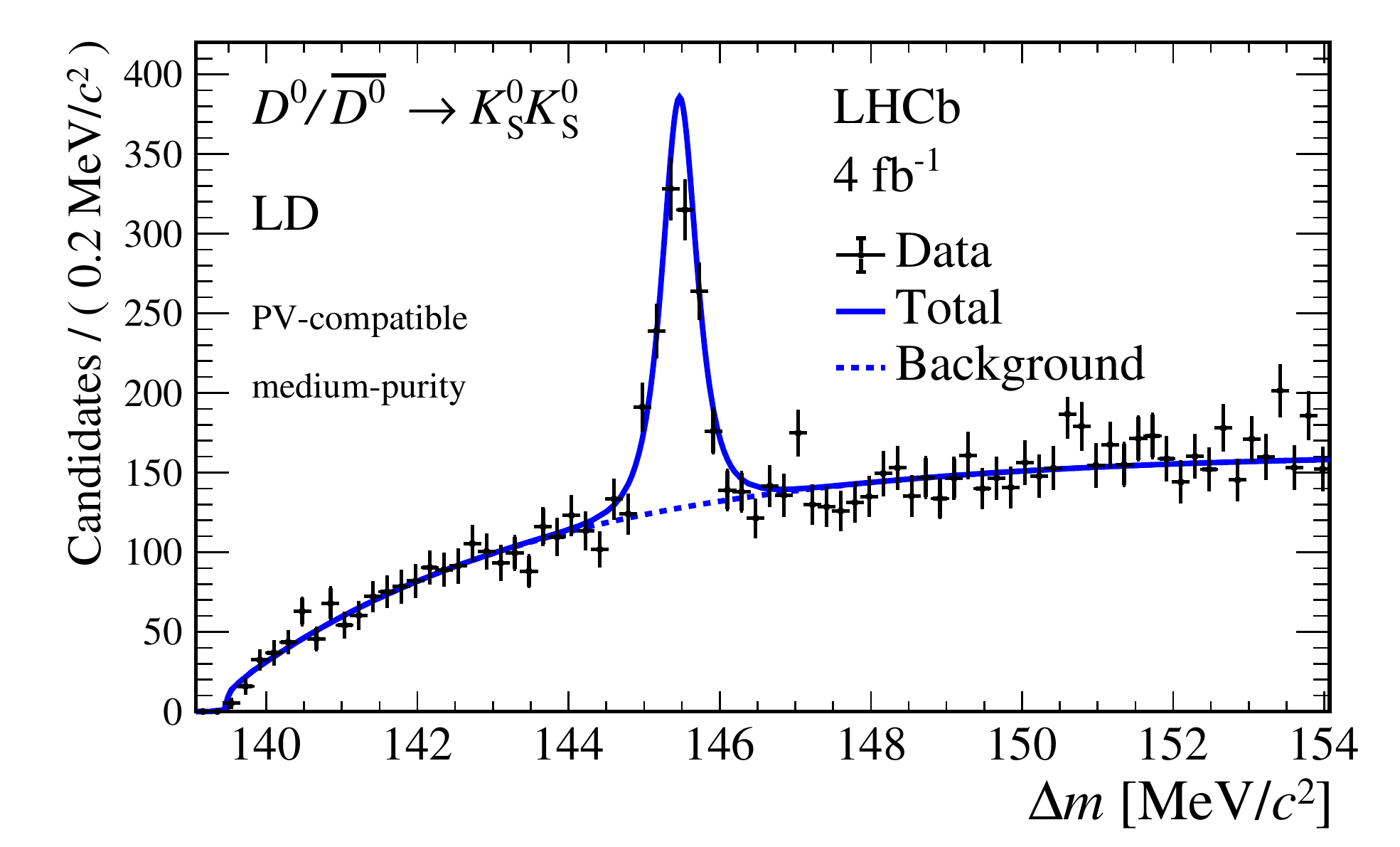}
\includegraphics[width=6.5 cm]{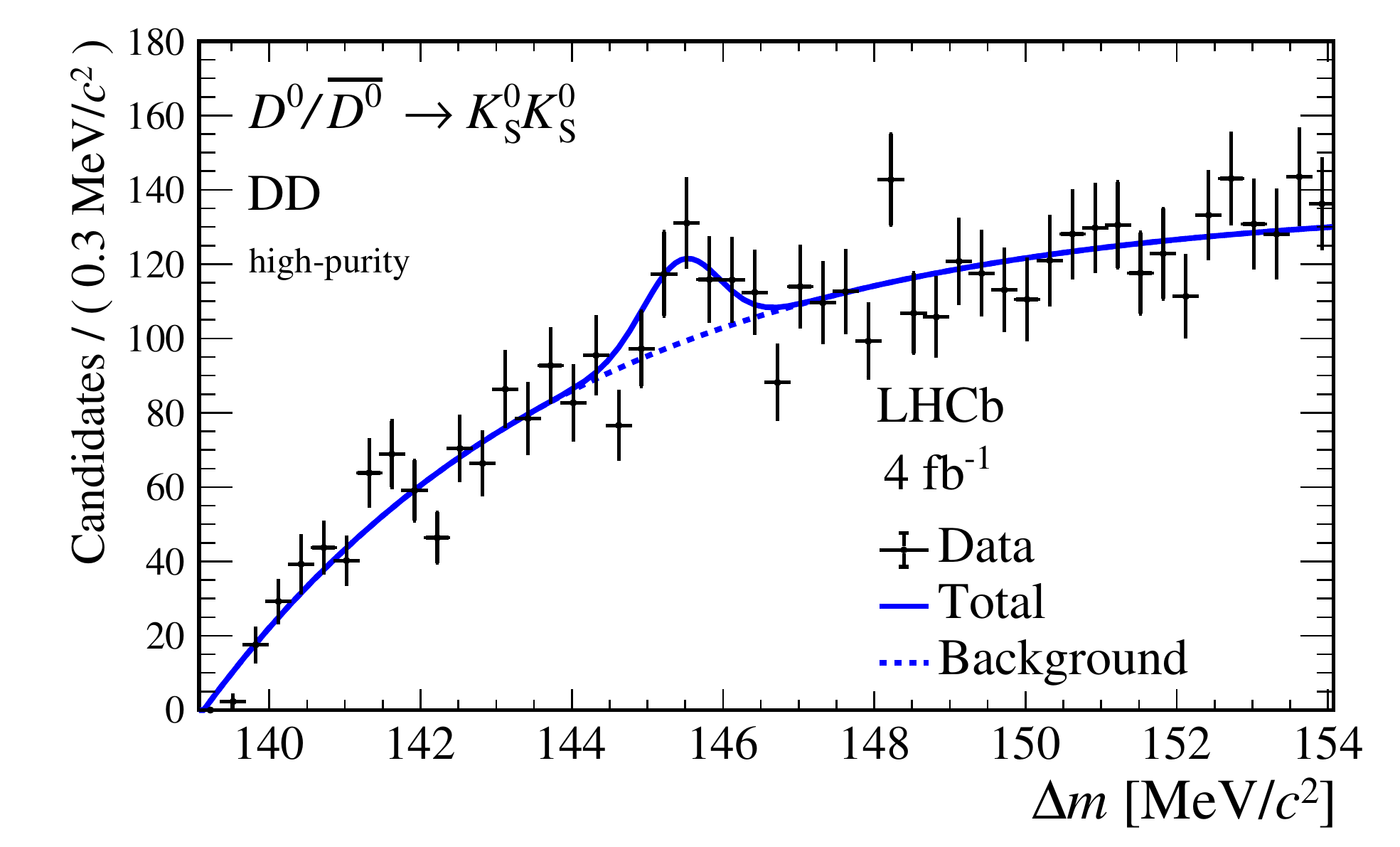}
\caption{Distributions and fit projections of the $\Delta m$ distribution for representative candidate categories (2017--2018 data~sample).}
\label{fit_KSKS}
\end{figure}

In order to measure the raw asymmetry of the signal, an~unbinned maximum-likelihood fit is performed to the joint distribution of the invariant masses of the $K_s^0$ mesons and the difference $\Delta m \equiv m(K_s^0 K_s^0 \pi^+) - m(K_s^0 K_s^0)$, simultaneously to candidates of both flavours.
The total probability density function is parameterised by the sum of the signal component, peaking in the three observables, and~seven additional components, which describe different kinds of background.
Fit projections on $\Delta m$ distribution are reported in Figure~\ref{fit_KSKS}, which shows how the mass resolution and the signal purity change between the different LL, LD and DD categories and between PV-compatible and PV-incompatible~\mbox{categories}.

A data sample of reconstructed $D^{*+} \to D^0 (\to K^+ K^-) \pi^+$ decays, whose $C\!P$ asymmetry is known to a precision much higher than that of the signal decay~\cite{LHCb:2016nxk}, is used as a calibration sample to correct for $D^{*+}$ production asymmetry and $\pi^+$ detection asymmetry.
As  experimental  asymmetries  are  a  function  of pion and $D^{*+}$ kinematics,  the~ kinematics  of the calibration samples are weighted to match those of the signal by using a multivariate classifier.
Various contributions to the systematic uncertainty are evaluated, namely related to weighting procedure, difference between signal and calibration modes and fit~model.

All the results obtained in each subsample are statistically compatible with each other and their combination gives:
\begin{equation*}
    \mathcal{A}_{C\!P}(D^0 \to K_s^0 K_s^0) = (-3.1 \pm 1.2 \pm 0.4 \pm 0.2)\%,
\end{equation*}
where the first uncertainty is statistical, the~second systematic and the third is related to the knowledge of $\mathcal{A}_{C\!P}(D^0 \to K^+ K^-)$~\cite{Aaij:2021uvv}.
This measurement, the~most precise of this quantity to date, is in agreement with the previous determinations~\cite{Bonvicini:2000qm,Dash:2017heu,Aaij:2015fua} and is compatible with no $C\!P$ violation at the level of 2.4 standard~deviations.

\subsection{Search for $C\!P$ violation in $D_{(s)}^+ \to h^+ \pi^0$ and $D_{(s)}^+ \to h^+ \eta$ decays}

The study of $C\!P$ asymmetry in the two-body $D_{(s)}^+ \to h^+ \pi^0$ and $D_{(s)}^+ \to h^+ \eta$ decays provides interesting tests of the SM.
In particular, SCS $D_s^+ \to K^+ \pi^0$, $D^+ \to \pi^+ \eta$ and $D_s^+ \to K^+ \eta$ decays receive contributions from two different weak phases allowing $C\!P$ violation at tree-level, expected to be of the order of $10^{-3}$--$10^{-4}$ according to the SM~\cite{Cheng:2012wr}.
In the SM, the~$C\!P$ asymmetry of the SCS $D^+ \to \pi^+ \pi^0$ mode is expected to be zero as a result of isospin constraints~\cite{Pirtskhalava:2011va,Buccella:2019kpn,Bhattacharya:2012ah,Bause:2020obd}.
It is possible to define the isospin sum rule
\begin{equation}
       R = \frac{\mathcal{A}_{C\!P} (D^0 \to \pi^+ \pi^-)}{1 + \frac{\tau_{D^0}}{\mathcal{B}_{+-}}\left(\frac{\mathcal{B}_{00}}{\tau_{D^0}} + \frac{2}{3}\frac{\mathcal{B}_{+0}}{\tau_{D^+}}\right) }
           + 
           \frac{\mathcal{A}_{C\!P} (D^0 \to \pi^0 \pi^0)}{1 + \frac{\tau_{D^0}}{\mathcal{B}_{00}}\left(\frac{\mathcal{B}_{+-}}{\tau_{D^0}} + \frac{2}{3}\frac{\mathcal{B}_{+0}}{\tau_{D^+}}\right) }
           - 
           \frac{\mathcal{A}_{C\!P} (D^+ \to \pi^+ \pi^0)}{1 + \frac{3}{2}\frac{\tau_{D^+}}{\mathcal{B}_{+0}}\left(\frac{\mathcal{B}_{00}}{\tau_{D^0}} + \frac{\mathcal{B}_{+-}}{\tau_{D^0}}\right) }\;,
\label{eq:sumrule}
\end{equation}
where $\tau_D^+$ and $\tau_D^0$ represent the $D^+$ and $D^0$ lifetimes and $\mathcal{B}_{+-}$, $\mathcal{B}_{00}$ and $\mathcal{B}_{+0}$ are the branching fractions of the $D^0 \to \pi^+ \pi^-$, $D^0 \to \pi^0 \pi^0$ and $D^+ \to \pi^+ \pi^0$ decays, respectively.
A non-zero value of $\mathcal{A}_{C\!P}(D^+ \to \pi^+ \pi^0)$, associated with a verification that $R$ is consistent with 0, would represent a signal of new~physics.

The Belle collaboration has reported the measurement of $\mathcal{A}_{C\!P}(D^+ \to \pi^+ \pi^0) = (2.31 \pm 1.24 \pm0.23)\%$, corresponding to $R = (-2.2 + 2.7)\times 10^{-3}$~\cite{Babu:2017bjn}, and, more recently, precise measurements of $\mathcal{A}_{C\!P} (D_s^+ \to K^+ \pi^0)$, $\mathcal{A}_{C\!P} (D_s^+ \to \pi^+ \eta)$ and $\mathcal{A}_{C\!P} (D_s^+ \to K^+ \eta)$ with uncertainties ranging from 0.4\% to 4.5\%~\cite{Guan:2021xer}.
The measurements of the $C\!P$ asymmetries in the $D_{(s)}^+ \to h^+ \pi^0$ and $D_{(s)}^+ \to h^+ \eta$ decays (except for $D_s^+ \to \pi^+ \pi^0$, that proceeds via an annihilation topology decay and is therefore highly suppressed) performed by LHCb with an integrated luminosity of $9~\mathrm{fb}^{-1}$ and $6~\mathrm{fb}^{-1}$, respectively, are summarised~here.

The $\pi^0$ and $\eta$ mesons are reconstructed in the $e^+ e^- \gamma$ final state, which allows the secondary decay vertex to be reconstructed, suppressing background originating from the $pp$ collisions.
This final state receives contributions from the $\pi^0 \to \gamma \gamma$ and $\eta \to \gamma \gamma$ decays, where one photon is converted to an $e^+ e^-$ pair after interacting with the detector material, and~the suppressed Dalitz $\pi^0 \to e^+ e^- \gamma$ and $\eta \to e^+ e^- \gamma$ decays.
During the signal reconstruction process, a~bremsstrahlung-recovery algorithm associates additional deposits from soft photons to those produced by the electrons in the electromagnetic calorimeter.
Decays with a total of either zero or one bremsstrahlung photon per $e^+ e^-$ pair are selected.
After the full selection is applied, about 86\% of signal decays are found to be due to photon conversions, and~the remaining part to the Dalitz~decays.

\begin{figure}[h]
\centering
\includegraphics[width=6.5 cm]{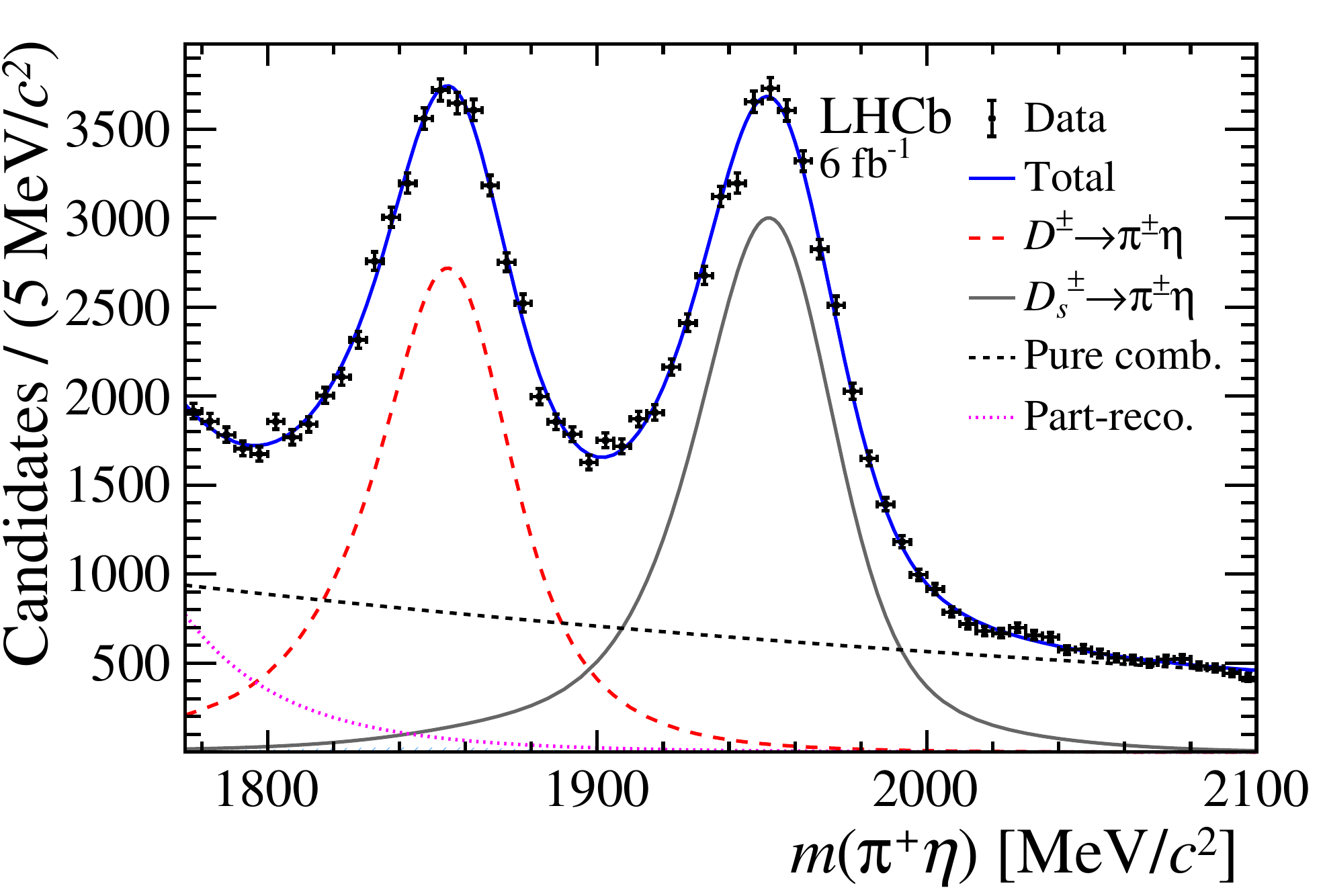}
\includegraphics[width=6.5 cm]{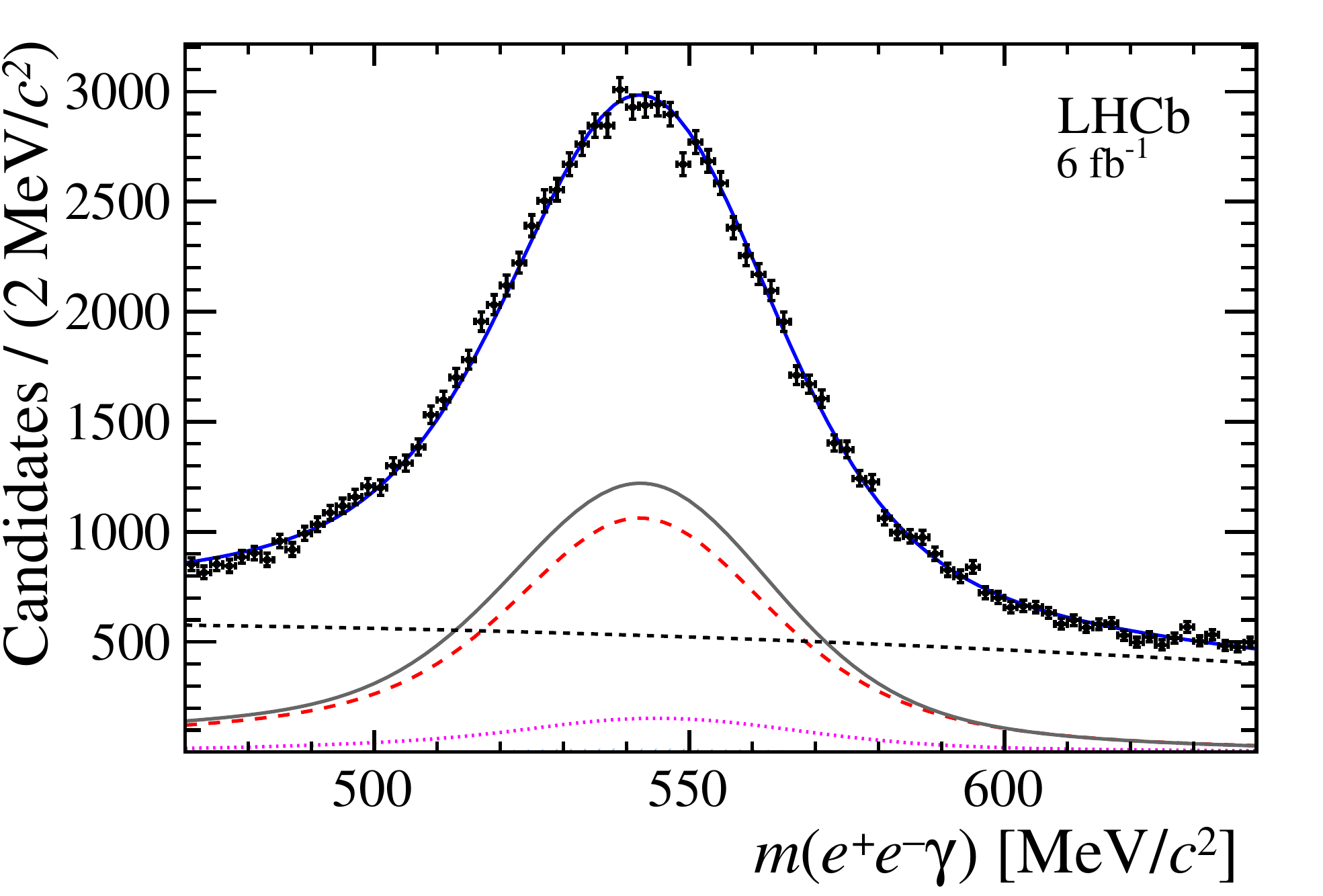}
\includegraphics[width=6.5 cm]{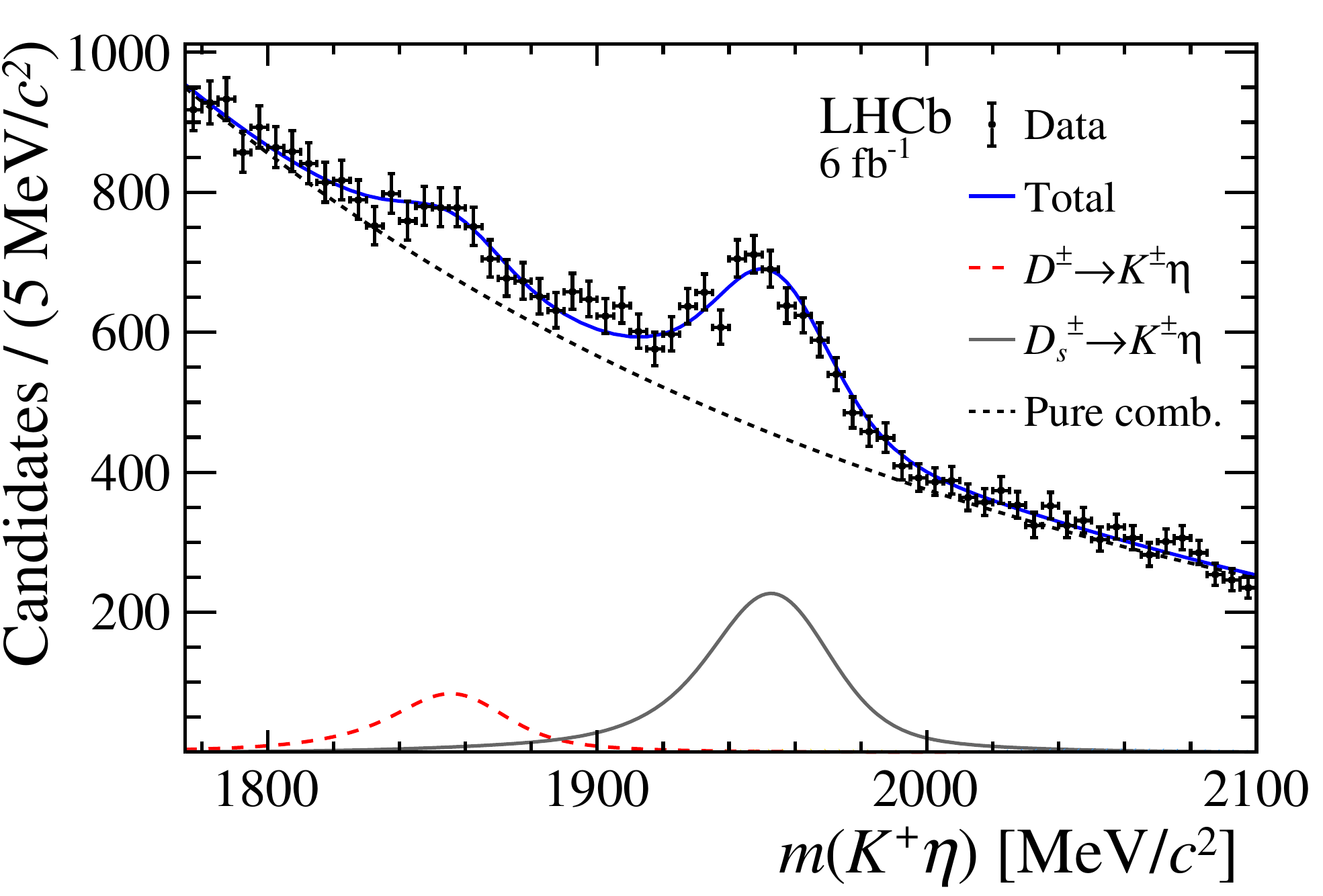}
\includegraphics[width=6.5 cm]{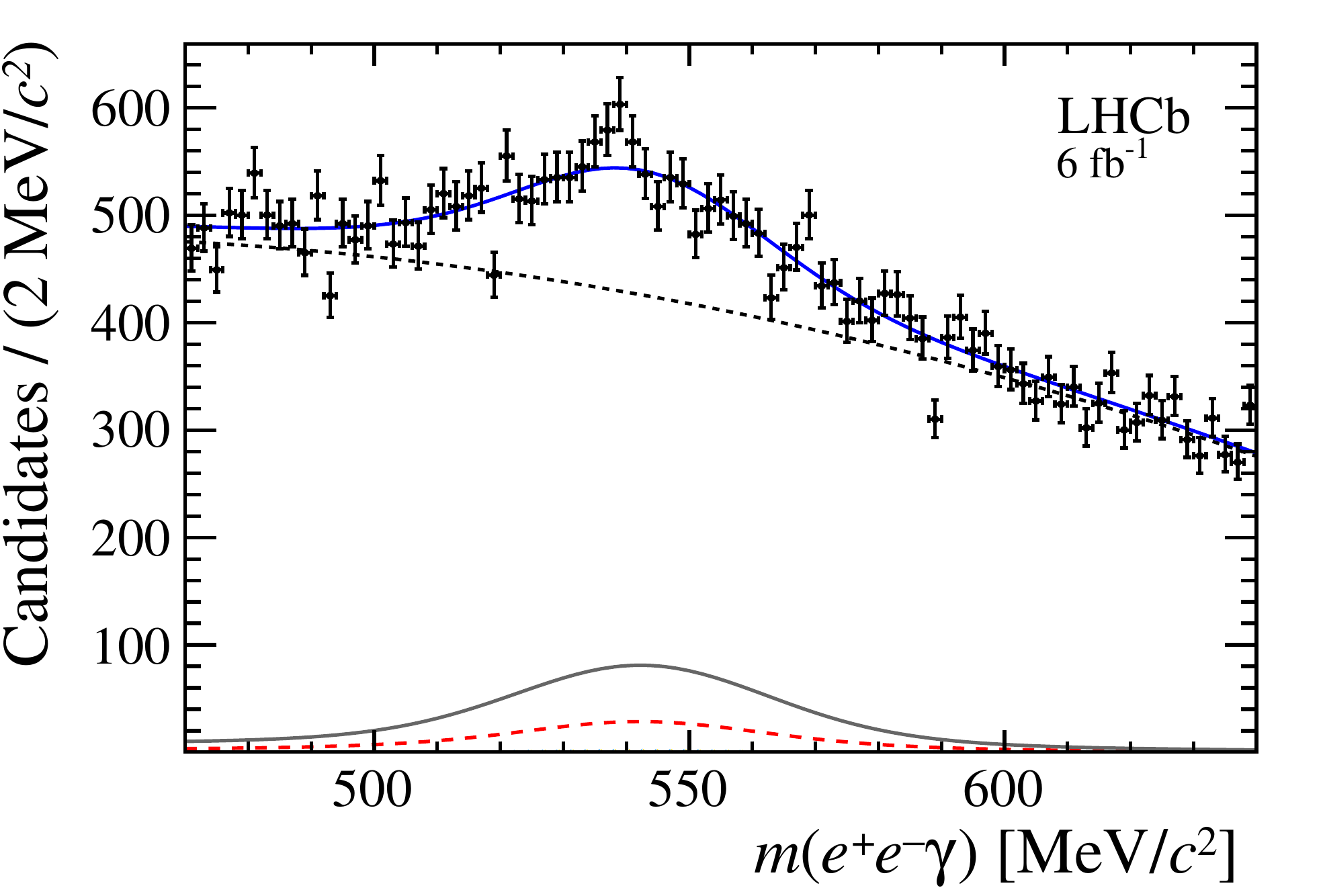}
\caption{Distribution of (\textbf{left}) $m(h^+ \eta)$ and (\textbf{right}) $m(e^+ e^- \gamma)$ for (\textbf{top}) $D_{(s)}^+ \to \pi^+ \eta$ and (\textbf{bottom}) $D_{(s)}^+ \to K^+ \eta$ candidates, summed over all categories of the simultaneous fit, with~projections of the fit result overlaid. $D^+ \to h^+ \eta$ contribution is shown in dashed red, $D^+_s \to h^+ \eta$ in solid grey, pure combinatorial decays in dashed black and partially reconstructed background in dotted magenta. The~misidentification background is too small to be~seen.}
\label{fit_heta}
\end{figure}

\begin{figure}[h]
\centering
\includegraphics[width=6.5 cm]{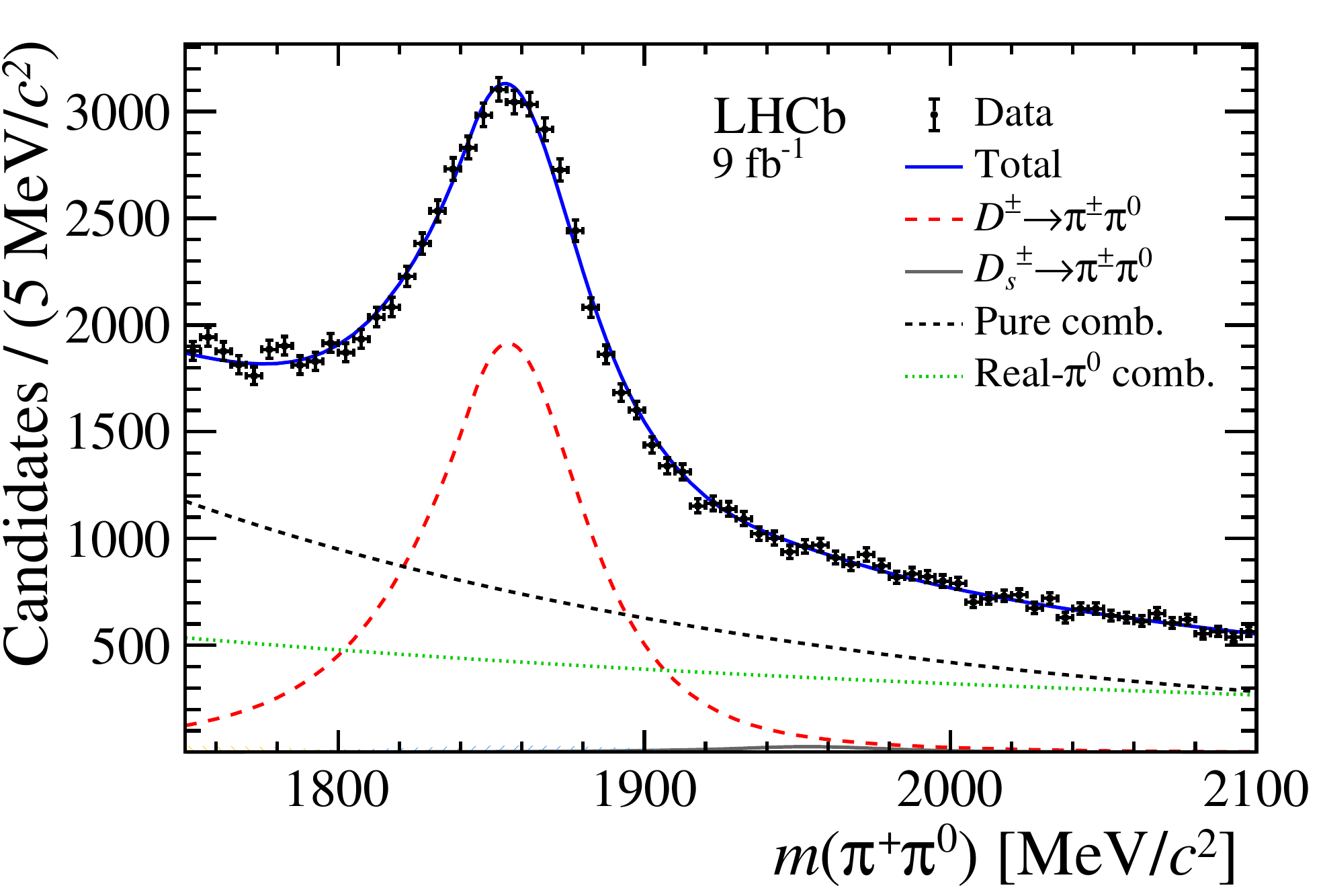}
\includegraphics[width=6.5 cm]{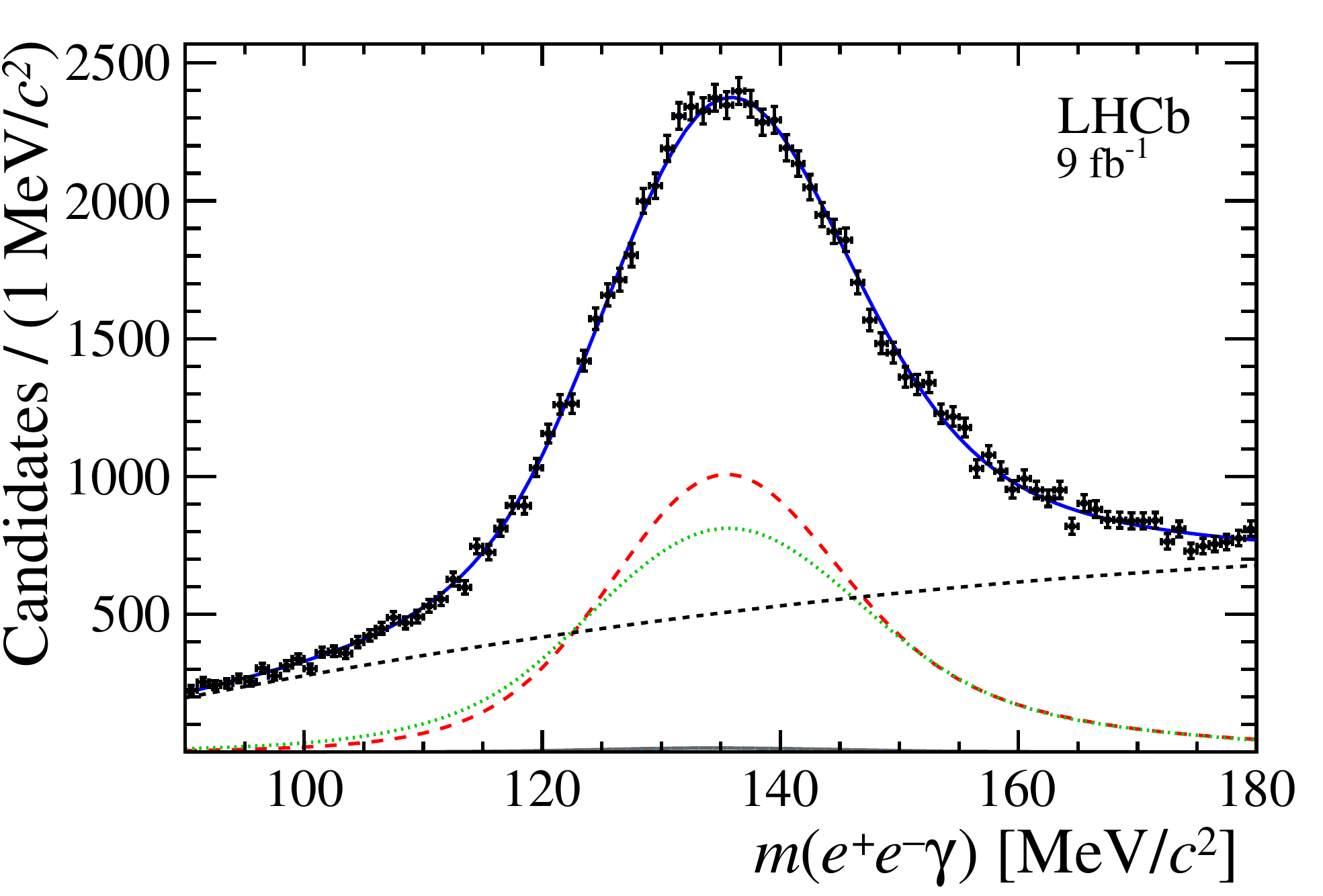}
\includegraphics[width=6.5 cm]{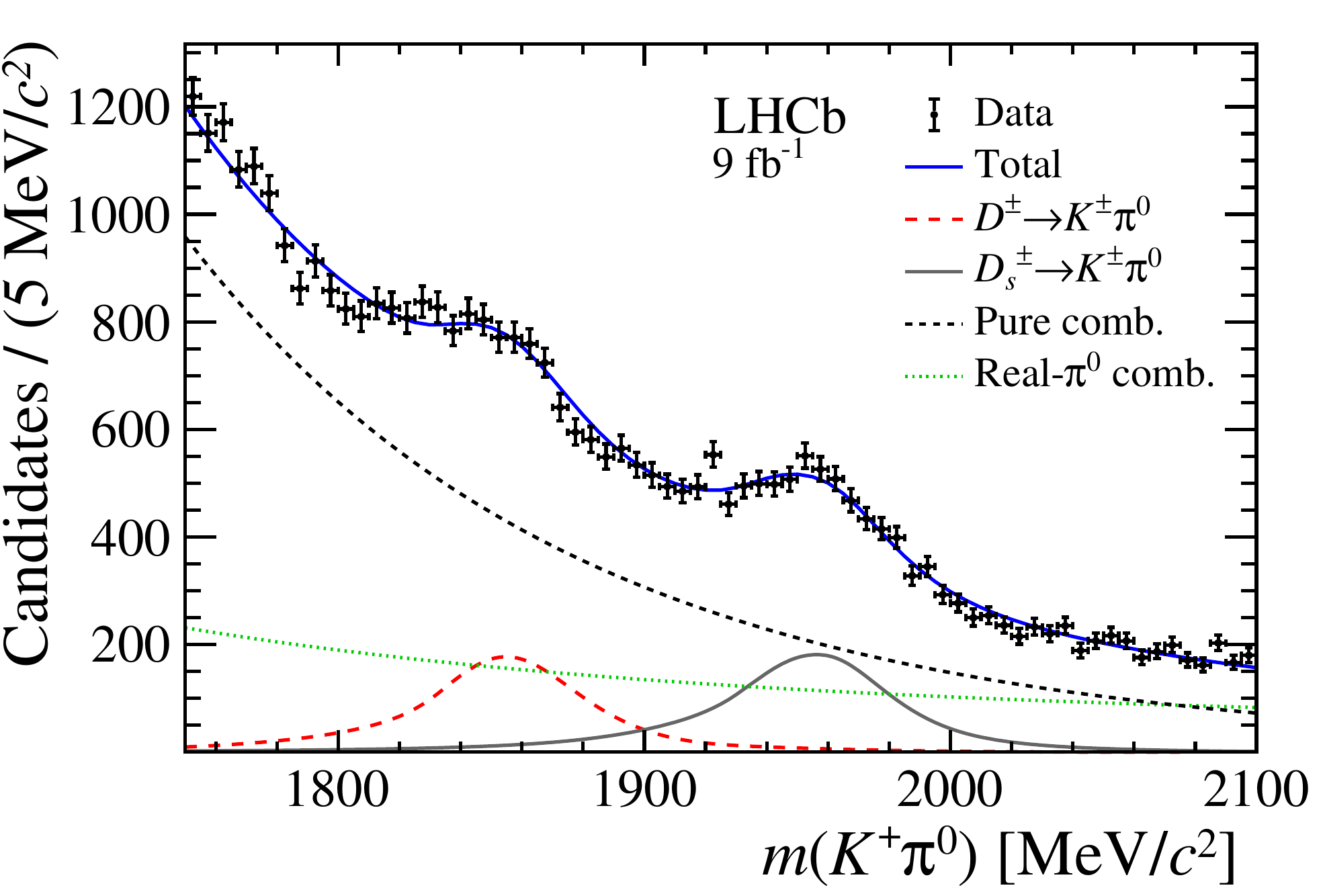}
\includegraphics[width=6.5 cm]{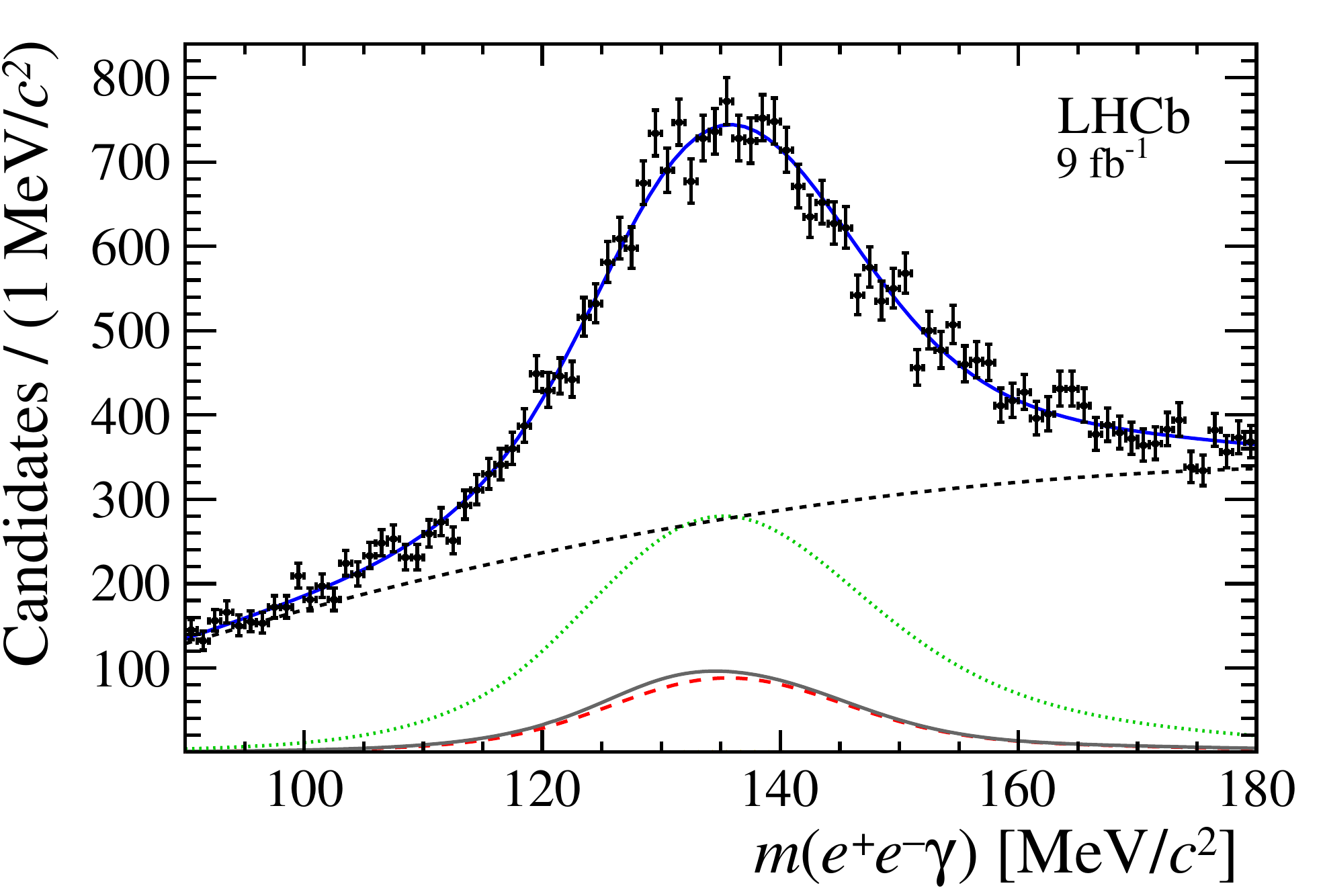}
\caption{Distribution of (\textbf{left}) $m(h^+ \pi^0)$ and (\textbf{right}) $m(e^+ e^- \gamma)$ for (\textbf{top}) $D_{(s)}^+ \to \pi^+ \pi^0$ and (\textbf{bottom}) $D_{(s)}^+ \to K^+ \pi^0$ candidates, summed over all categories of the simultaneous fit, with~projections of the fit result overlaid. $D^+ \to h^+ \pi^0$ contribution is shown in dashed red, $D^+_s \to h^+ \pi^0$ in solid grey, pure combinatorial decays in dashed black and real-$\pi^0$ combinatorial background in dotted green. The~misidentification background is too small to be~seen.}
\label{fit_hpi0}
\end{figure}

The raw asymmetries are measured by means of two-dimensional extended simultaneous unbinned maximum-likelihood fits to the invariant mass $m(e^+ e^- \gamma)$ and the invariant mass difference $m(h^+ h^0) \equiv m(h^+ e^+ e^- \gamma) - m(e^+ e^- \gamma) + M(h^0)$, where $M(h^0)$ is the known $\pi^0$ and $\eta$ mass~\cite{Zyla:2020zbs}.
Two-dimensional probability density functions are used to model different contributions, namely signal decays, pure combinatorial background, combinatorial background due to real $\pi^0$ meson combined with an unrelated track, misidentification background and partially reconstructed low-mass background.
The fits are performed simultaneously in categories depending on the running period, the~number of recovered bremsstrahlung photons, the~$h^+$ type (pion or kaon) and the candidate charge.
The results of the fits are shown in Figures~\ref{fit_heta} and \ref{fit_hpi0}.

In order to cancel the production and detection asymmetries and obtain the $C\!P$ asymmetries, a~calibration sample of $D_{(s)}^+ \to K_s^0 h^+$ decays is used, after~weighting its kinematic distributions to match those of the signal candidates.
The values of $\mathcal{A}_{C\!P}(D_{(s)}^+ \to K_s^0 h^+)$ are known with sub-percent precision~\cite{Aaij:2019vnt}, significantly higher than the precision of this measurement. 
Systematic uncertainties related to the choice of the fit model, presence of $D_{(s)}^+$ candidates produced in secondary decays, different selection between signal and calibration sample, kinematic weighting and uncertainties of $C\!P$ asymmetries of the calibration decays are~evaluated.

The results are
\begin{alignat*}{7}
       \mathcal{A}_{C\!P}(D^+ \to \pi^+\pi^0) 	&= (-&&1.3 &&\pm 0.9 &&\pm 0.6 &)\%, \\
       \mathcal{A}_{C\!P}(D^+ \to K^+\pi^0) 	&= (-&&3.2 &&\pm 4.7 &&\pm 2.1 &)\%, \\
       \mathcal{A}_{C\!P}(D^+ \to \pi^+\eta)   &= (-&&0.2 &&\pm 0.8 &&\pm 0.4 &)\%, \\
       \mathcal{A}_{C\!P}(D^+ \to K^+\eta) 	&= (-&&6 &&\pm 10 &&\pm 4 &)\%, \\
       \mathcal{A}_{C\!P}(D^+_s \to K^+\pi^0) 	&= (-&&0.8 &&\pm 3.9 &&\pm 1.2 &)\%, \\
       \mathcal{A}_{C\!P}(D^+_s \to \pi^+\eta)  &= (&&0.8 &&\pm 0.7 &&\pm 0.5 &)\%, \\
       \mathcal{A}_{C\!P}(D^+_s \to K^+\eta)   &= (&&0.9 &&\pm 3.7 &&\pm 1.1 &)\%,
\end{alignat*}
where the first uncertainty is statistical and the second systematic~\cite{Aaij:2021wfw}.
All of the results are consistent with previous determinations~\cite{Babu:2017bjn,Guan:2021xer} and with no $C\!P$ violation, and~the first five constitute the most precise measurements to date of those observables.
An average of $\mathcal{A}_{C\!P}(D^+ \to \pi^+ \pi^0)$ between this result, the~measurement by Belle~\cite{Babu:2017bjn} and a measurement by CLEO~\cite{Mendez:2009aa} gives $\mathcal{A}_{C\!P}(D^+ \to \pi^+ \pi^0) = (0.43 \pm 0.79)\%$, corresponding to a value of ${R = (0.1 \pm 2.4)\times 10^{-3}}$, consistent with~0.

\subsection{Search for time-dependent $C\!P$ violation in $D^0 \to K^+ K^-$ and $D^0 \to \pi^+ \pi^-$ decays}

The magnitude of the $\Delta Y_f$ parameter is expected to be about $2 \times 10^{-5}$~\cite{Kagan:2020vri,Li:2020xrz}.
The $\Delta Y_f$ parameter is approximately equal, up~to 1\% relative corrections ({For a detailed description of the theoretical parametrisation of mixing in $D^0$ decays, see Refs.~\cite{Kagan:2020vri,Pajero:2021jev,Pajero:2021dgd}}), to~the negative of the parameter $A_\Gamma^f$ that has been measured by the BaBar~\cite{Lees:2012qh}, CDF~\cite{Aaltonen:2014efa}, Belle~\cite{Staric:2015sta} and LHCb~\cite{Aaij:2015yda,Aaij:2017idz,Aaij:2019yas} collaborations, and~the world average, assuming no difference is present between the $K^+ K^-$ and $\pi^+ \pi^-$ final states, is $\Delta Y = (3.1 \pm 2.0) \times 10^{-4}$~\cite{Amhis:2019ckw}.
The measurement of $\Delta Y_f$ performed by LHCb with data collected during the LHC Run~2, corresponding to an integrated luminosity of $6~\mathrm{fb}^{-1}$, using $D^{*+} \to D^0 \pi^+$ decays originated from primary $pp$ interactions, is reported~here.

\begin{figure}[h]
\centering
\includegraphics[width=4.3 cm]{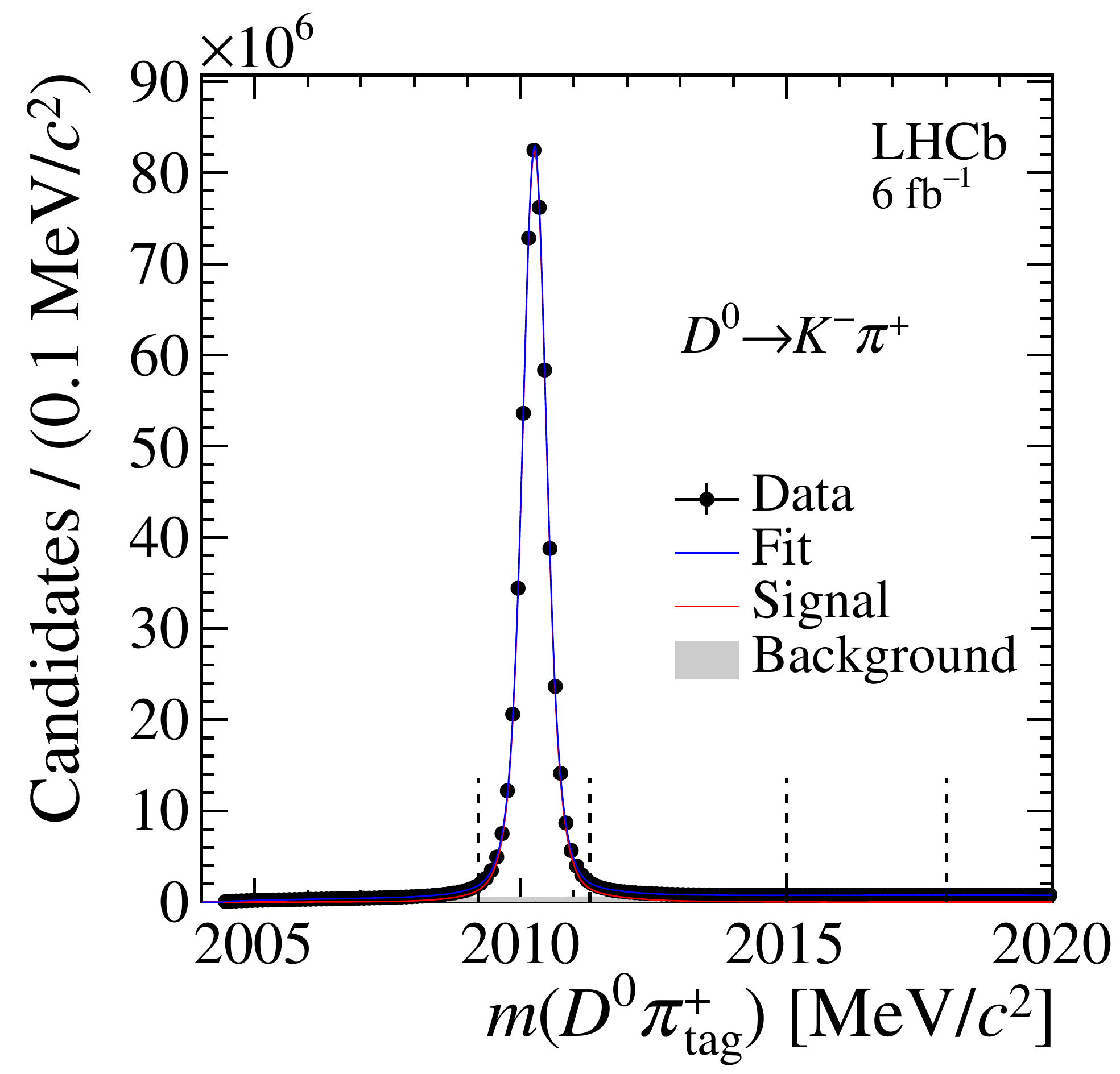}
\includegraphics[width=4.3 cm]{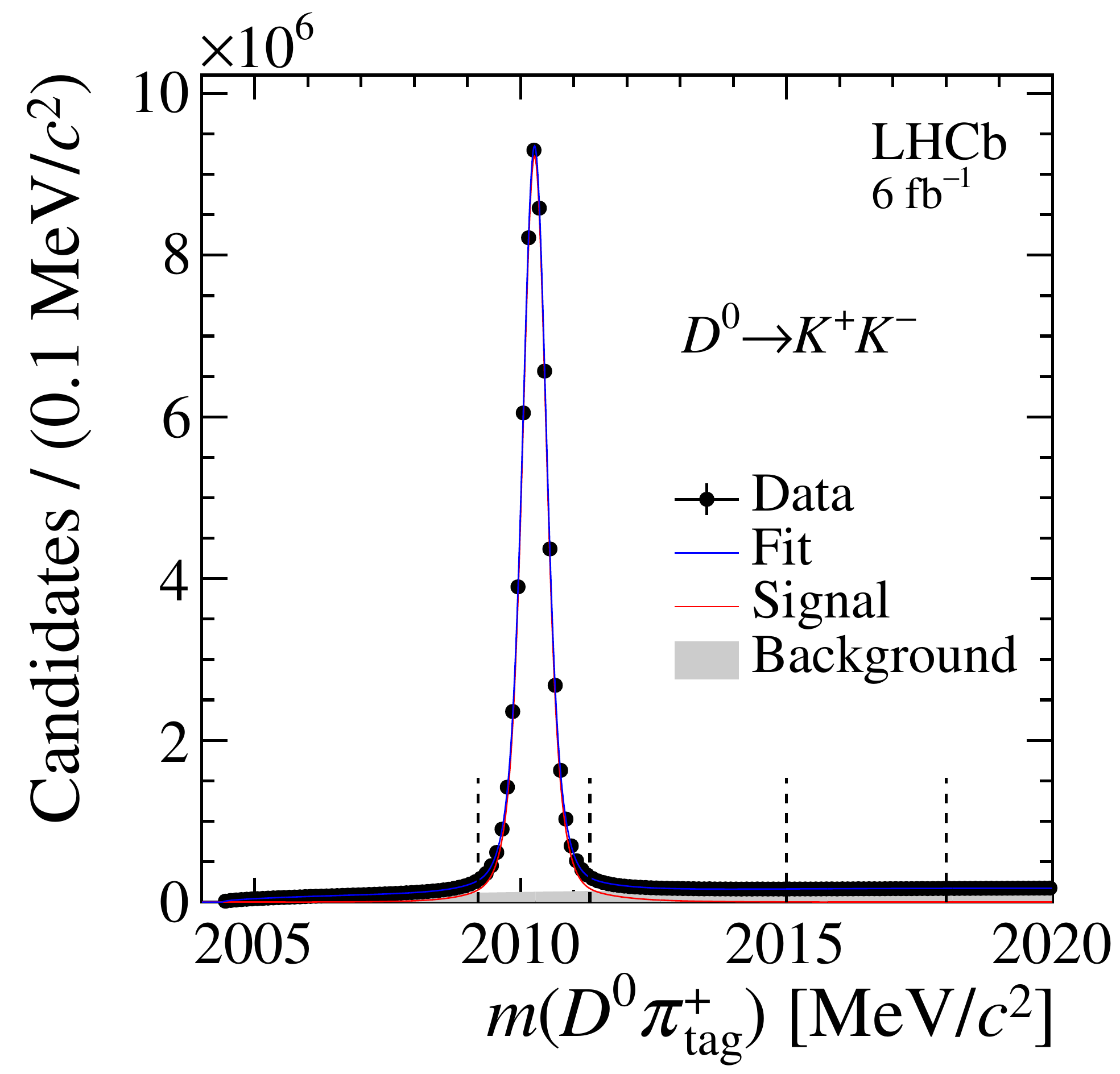}
\includegraphics[width=4.3 cm]{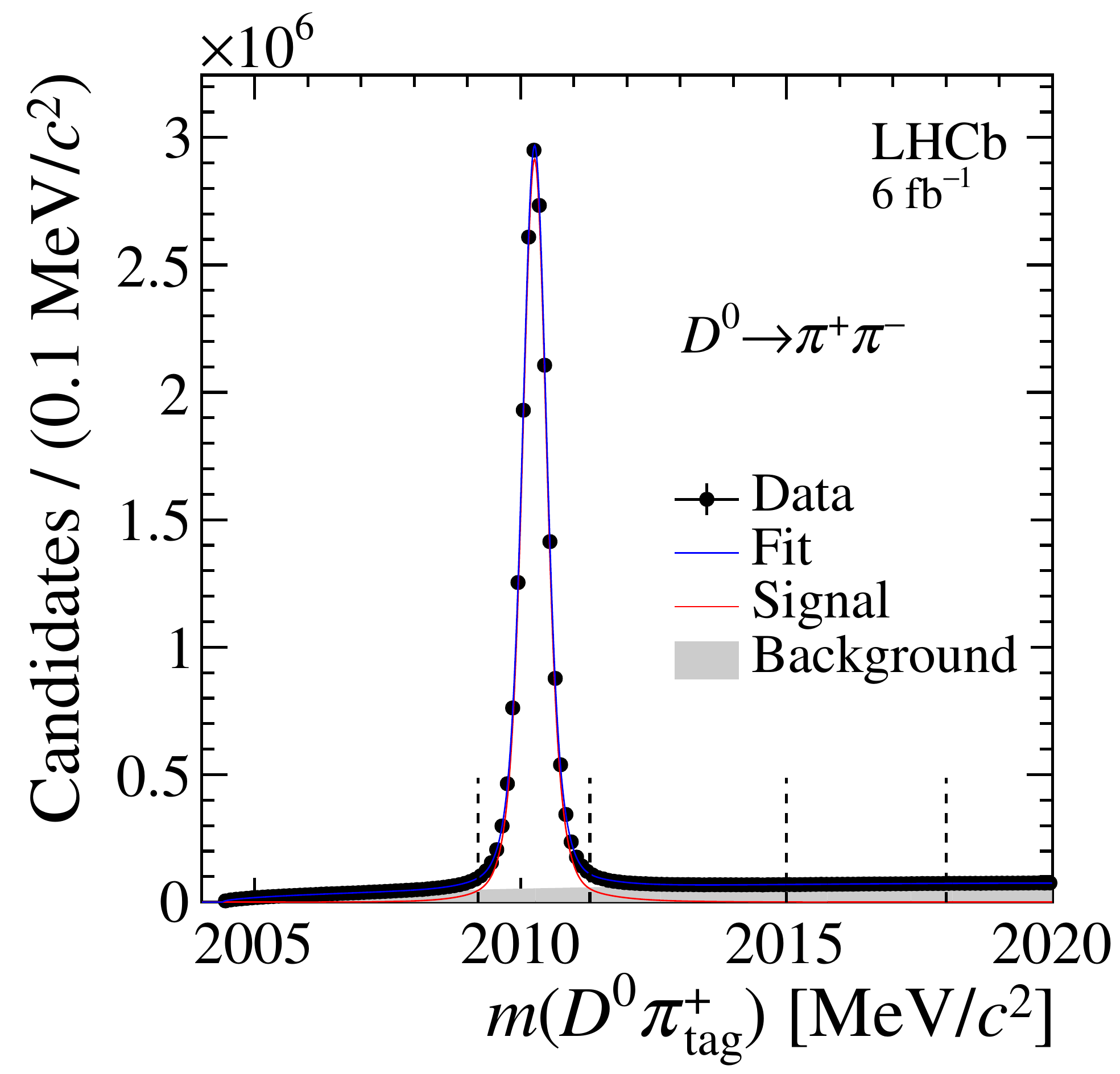}
\caption{Distribution of $m(D^0 \pi^+)$ for (\textbf{left}) $D^0 \to K^- \pi^+$, (\textbf{center}) $D^0 \to K^+ K^-$ and (\textbf{right}) $D^0 \to \pi^+ \pi^-$ candidates. The~signal window and the lateral window used to remove the combinatorial background (grey filled area) are delimited by the vertical dashed lines. Fit projections are~overlaid.}
\label{fit_mass_DY}
\end{figure}

Three decay modes are reconstructed: $D^0 \to K^+ K^-$, $D^0 \to \pi^+ \pi^-$ and $D^0 \to K^- \pi^+$, where the $D^0$ meson is associated with the accompanying pion in the $D^{*+} \to D^0 \pi^+$ decay to determine its flavour when produced.
The $D^0 \to K^- \pi^+$ mode has the same topology and kinematic distributions very similar to those of the signal channels and its $C\!P$ asymmetry is known to be smaller than the current experimental uncertainty.
For these reasons, the~$D^0 \to K^- \pi^+$ mode is used as a control sample to develop and validate the analysis.
The data sample is divided in 21 intervals of $D^0$ decay times in the range of (0.45--8) $\tau_{D^0}$, where $\tau_{D^0}$ is the nominal $D^0$ lifetime~\cite{Zyla:2020zbs}.
In each decay time interval the signal yield is obtained by means of background subtraction in the $m(D^0 \pi^+)$ distribution, where the signal window is defined as [2009.2, 2011.3]~{MeV}/$c^2$ and the background candidates are taken from the lateral window [2015, 2018]~{MeV}/$c^2$.
The weight assigned to the background candidates is determined with a binned maximum-likelihood fit to the $m(D^0 \pi^+)$ distribution.
The fit relies on an empirical model to describe the signal and the combinatorial background contributions.
The $m(D^0 \pi^+)$ distributions and the fit results are shown in Figure~\ref{fit_mass_DY}.
After the full selection and background subtraction, the~number of candidates in the signal region are 519, 58 and 18 millions for the $D^0 \to K^- \pi^+$, $D^0 \to K^+ K^-$ and $D^0 \to \pi^+ \pi^-$ decay channels, respectively.
The candidates are split in subsamples according to year of data taking and magnet polarities. The~analysis is performed on each subsample and the results are combined at the~end.

\begin{figure}[h]
\centering
\includegraphics[width=6.5 cm]{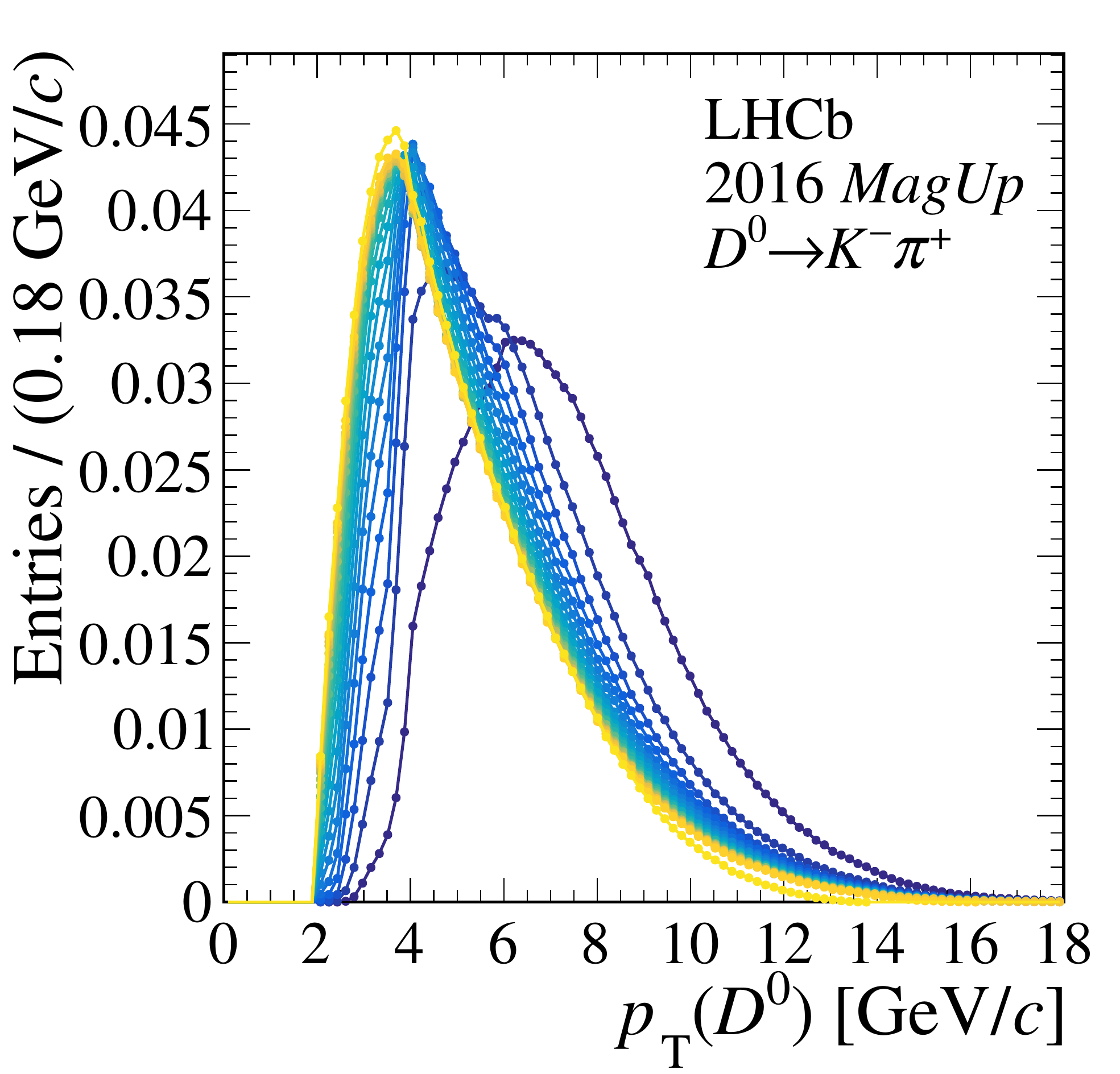}
\includegraphics[width=6.5 cm]{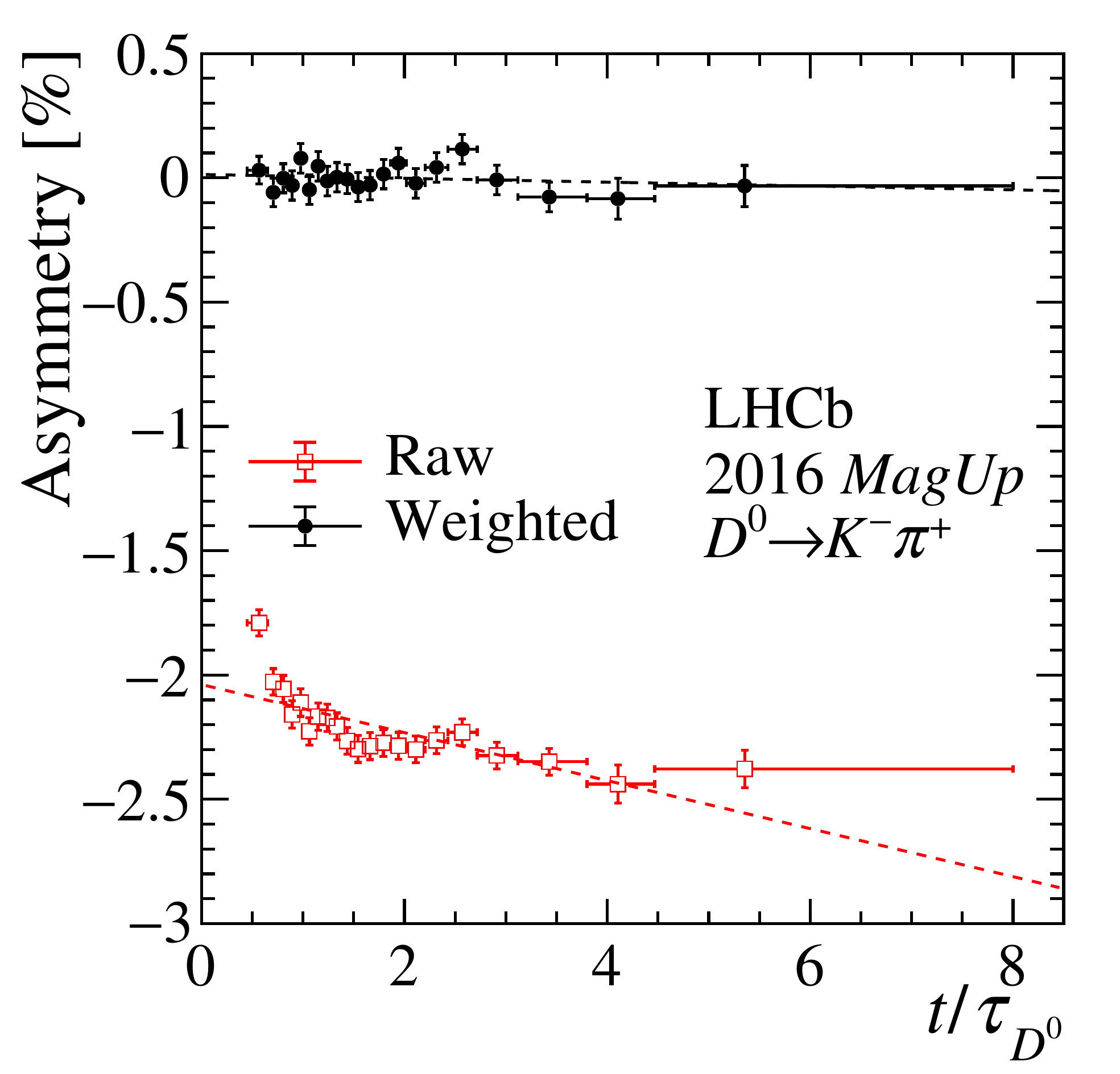}
\caption{(\textbf{Left}) Normalised distributions of the $D^0$ transverse momentum, in~different colours for each decay-time interval. Decay time increases from blue to yellow colour. (\textbf{Right}) Linear fit to the time-dependent asymmetry (red) before and (black) after the kinematic weighting. The~plots correspond to the $D^0 \to K^- \pi^+$ candidates recorded in 2016 with the magnet polarity pointing~upwards.}
\label{rew_Kpi_DY}
\end{figure}

In principle, the~measurement of $\Delta Y_f$ is insensitive to time-independent asymmetries such as the production and detection asymmetries, that depend only on the kinematics of the $D^{*+}$ meson and the tagging pion.
However, an~indirect time dependence of the nuisance asymmetries exists because of the correlation between the kinematic variables and the $D^0$ decay time introduced by the selection requirements.
The nuisance asymmetries are therefore removed by equalising the kinematics of tagging $\pi^+$ and $\pi^-$ and of $D^0$ and $\overline{D}^0$ candidates, by~weighting their kinematic distributions to their average.
Figure~\ref{rew_Kpi_DY} shows how the $D^0$ transverse momentum is correlated with the decay time and the effect of the weighting procedure to the time-dependent raw asymmetry.
A side effect of the kinematic weighting is that, because~of the correlation between the decay time and momentum of the $D^0$ meson, a~possible time-dependent asymmetry due to a non-zero value of $\Delta Y_f$ would be partially cancelled by the kinematic weighting, inducing a dilution in the measurement.
By introducing different artificial values of $\Delta Y_{K^-\pi^+}$ in the $D^0 \to K^- \pi^+$ sample, up~to values as large as 100 times the statistical uncertainty of the final measurement, the~dilution is found to have a linear effect on the measured value of $\Delta Y_{K^-\pi^+}$ equal to $(96.9 \pm 0.1)\%$ of the introduced value.
The results of all decay channels are therefore corrected to account for this dilution~factor.

The presence of $D^{*+}$ mesons originating from $B$ mesons in the selected sample can potentially induce a bias in the measured time-dependent asymmetry, due to a difference in production asymmetry between $D^{*+}$ mesons originating from secondary decays and $D^{*+}$ mesons produced in $pp$ collisions.
In addition, the~measurement of the decay time is performed with respect to the PV, resulting in a biased decay time towards larger values for secondary $D^{*+}$ mesons.
Although the secondary decays are suppressed by a requirement on the IP of the $D^0$ meson, some contamination is still present in the sample.
The size and asymmetry of this background is therefore assessed by means of a binned maximum-likelihood fit to the bidimensional distribution of IP and decay time of the $D^0 \to K^- \pi^+$ candidates, and~the effect of secondary decays on the final measurement is corrected for.
Systematic uncertainties are assessed related to background removal, mass difference between $D^{*+}$ and $D^{*-}$, correction due to secondary decays, presence of misidentified background peaking in $m(D^0 \pi^+)$ distribution and kinematic~weighting.

\begin{figure}[h]
\centering
\includegraphics[width=6.5 cm]{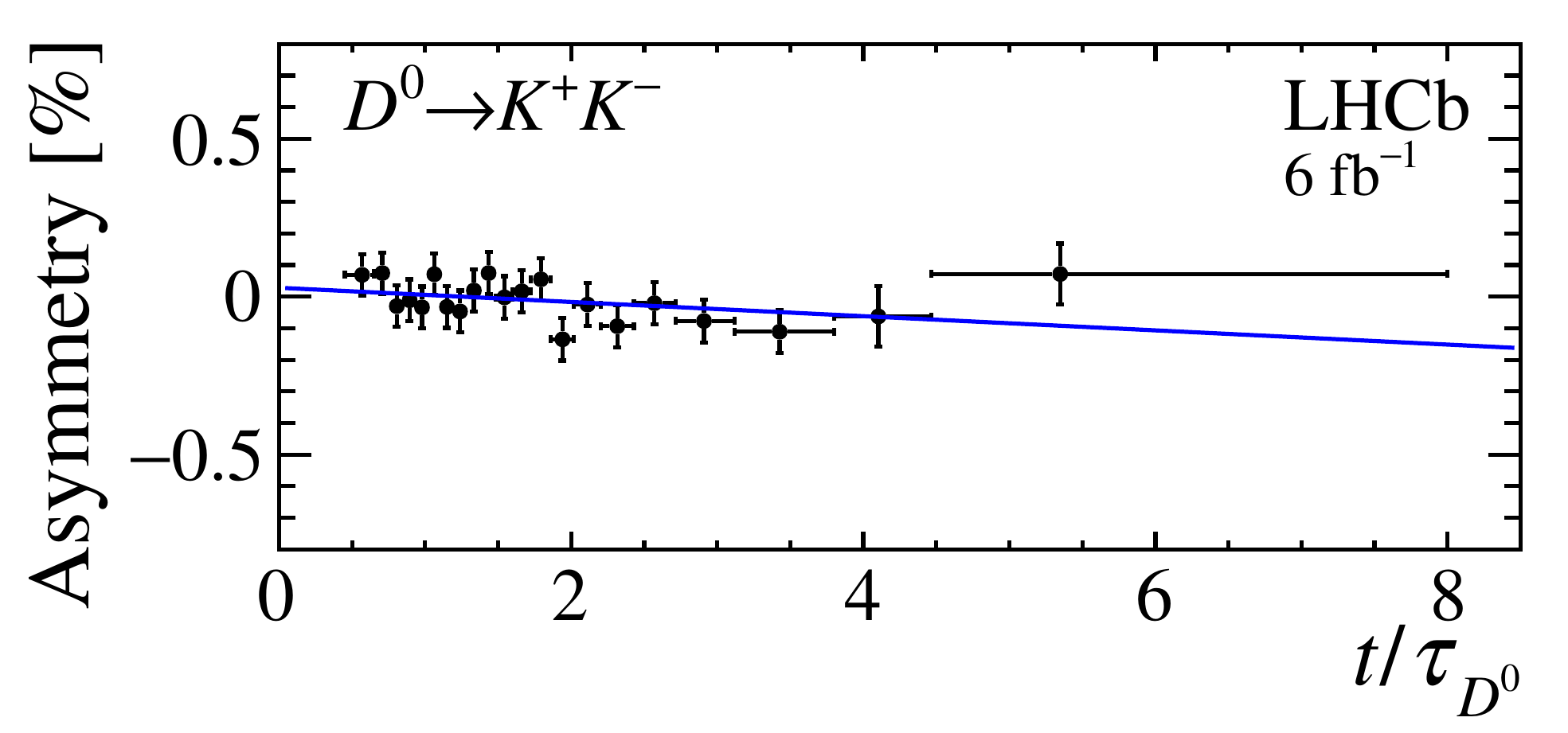}
\includegraphics[width=6.5 cm]{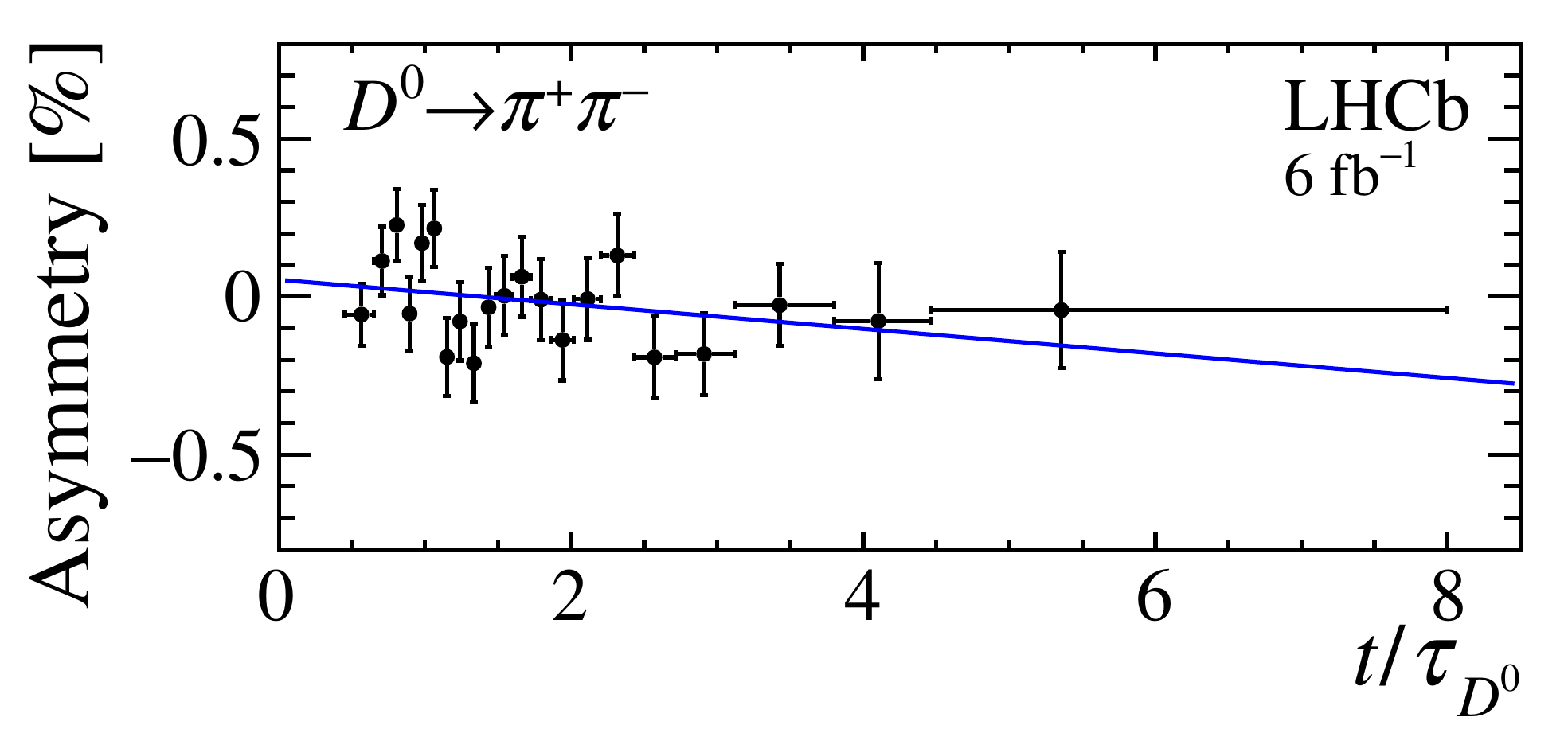}
\caption{Asymmetry as a function of $D^0$ decay time, for~(\textbf{left}) $D^0 \to K^+ K^-$ and (\textbf{right}) ${D^0 \to \pi^+ \pi^-}$ samples, with~linear fit~superimposed.}
\label{fit_KK_pipi_DY}
\end{figure}

The slope of the time-dependent asymmetry of the control sample is measured to be $(-0.4 \pm 0.5 \pm 0.2) \times 10^{-4}$, compatible with 0 as expected.
The time-dependent asymmetries of the $D^0 \to K^+ K^-$ and $D^0 \to \pi^+ \pi^-$ channels, after~the full selection, weighting and corrections, are displayed in Figure~\ref{fit_KK_pipi_DY}, with~linear fits superimposed.
The resulting slopes~are
\begin{align*}
    \Delta Y_{K^+ K^-} &= (-2.3 \pm 1.5 \pm 0.3) \times 10^{-4}, \\
    \Delta Y_{\pi^+ \pi^-} &= (-4.0 \pm 2.8 \pm 0.4) \times 10^{-4},
\end{align*}
where the first uncertainties are statistical and the second are systematic, compatible with each other and with previous determinations.
Assuming no final-state dependency, and~taking account of the correlation between the systematic uncertainties, the~combination of the two values gives
\begin{equation*}
    \Delta Y = (-2.7 \pm 1.3 \pm 0.3) \times 10^{-4},
\end{equation*}
consistent with zero within two standard deviations~\cite{Aaij:2021pyl}.
The combination with previous LHCb measurements~\cite{Aaij:2015yda,Aaij:2017idz,Aaij:2019yas} leads to
\begin{align*}
    \Delta Y_{K^+ K^-} &= (-0.3 \pm 1.3 \pm 0.3) \times 10^{-4}, \\
    \Delta Y_{\pi^+ \pi^-} &= (-3.6 \pm 2.4 \pm 0.4) \times 10^{-4}, \\
    \Delta Y &= (-1.0 \pm 1.1 \pm 0.3) \times 10^{-4}.
\end{align*}

This result, consistent with no time-dependent $C\!P$ violation, is the most precise measurement of $C\!P$ violation in charm decays and improves by nearly a factor of two the precision of the previous world average of $\Delta Y$~\cite{Amhis:2019ckw}.

\subsection{Observation of the mass difference between neutral charm-meson eigenstates}

There are currently no precise SM estimates of the mixing parameters of the $D^0$ meson.
The inclusive approach, that relies on the heavy-quark expansion, predicts a value of $y$ in large disagreement with experimental results~\cite{Carrasco:2014uya,Kirk:2017juj,Carrasco:2015pra,Bazavov:2017weg,Lenz:2020efu}.
The disagreement can be due to a violation of the quark-hadron duality or to a lifting of the GIM suppression for higher-dimension operators, even if these effects seem too small to explain the discrepancy~\cite{Jubb:2016mvq,Georgi:1992as,Ohl:1992sr,Bigi:2000wn,Bobrowski:2010xg,Bobrowski:2012jf}.
The exclusive approach, which calculates the mixing amplitudes at the hadron level by summing the contributions of all possible intermediate virtual states, predicts the size of $x$ and $y$ to be $\mathcal{O}(10^{-3})$~\cite{Falk:2001hx,Gronau:2012kq,Cheng:2010rv,Jiang:2017zwr,Wolfenstein:1985ft,Donoghue:1985hh}.

According to the current world averages, the~values of the mixing and $C\!P$-violating parameters of the $D^0$ meson are $x = (3.7 \pm 1.2)\times 10^{-3}$, $y = (6.8^{+0.6}_{-0.7})\times 10^{-3}$, $|q/p| = (0.951^{+0.053}_{-0.042})$ and $\phi = (-0.092^{+0.085}_{-0.079})$~\cite{Amhis:2019ckw}.
These parameters have been measured by studying mainly the $D^0 \to K^+ \pi^-$ decay, that allowed precise determinations of $y$ and the observation of mixing in the $D^0$ meson, and~the $D^0 \to K_s^0 \pi^+ \pi^-$ decay, which is particularly sensitive to $x$~\cite{Asner:2005sz,delAmoSanchez:2010xz,Peng:2014oda,Aaij:2015xoa,Aaij:2019jot}.
Here, the recent measurement of the mixing and $C\!P$ violation parameters in $D^0 \to K_s^0 \pi^+ \pi^-$ decays performed by the LHCb collaboration with the data collected between 2016 and 2018, corresponding to $5.4~\mathrm{fb}^{-1}$ of integrated luminosity, is~summarised.

\begin{figure}[h]
\centering
\includegraphics[width=8.5 cm]{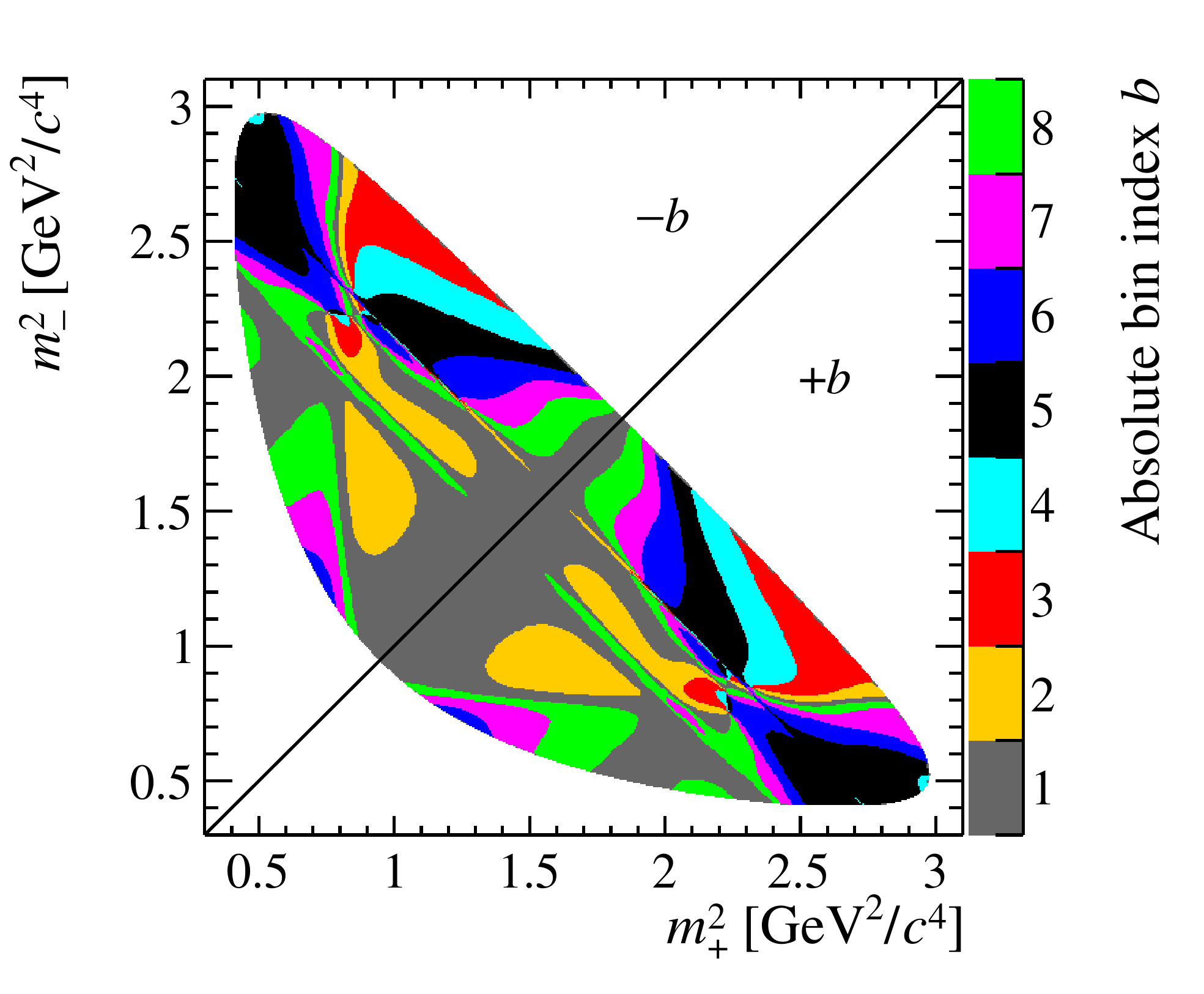}
\caption{Binning scheme of the $D^0 \to K_s^0 \pi^+ \pi^-$ Dalitz plot.}
\label{KSpipi_dalitz_bins}
\end{figure}

The analysis uses the ``bin-flip'' method~\cite{DiCanto:2018tsd}, that consists in measuring, as~a function of the $D^0$ decay time, the~ratio of the number of decays between symmetric bins in the Dalitz plot defined by the two squared invariant masses $m^2(K_s^0 \pi^\pm)$.
The ``bin-flip'' method is a model-independent approach and does not require an accurate modelling of the efficiency.
The Dalitz bins, illustrated in Figure~\ref{KSpipi_dalitz_bins}, are defined in such a way that the strong-phase difference between the $D^0$ and $\overline{D}^0$ amplitudes within each bin is nearly constant~\cite{Libby:2010nu}.
The set of bins for which $m^2_{+} > m^2_{-}$, where $m^2_{\pm}$ denotes $m^2(K_s^0 \pi^\pm)$, is given a positive index $+b$, whereas the other set, symmetric about the $m^2_{+} = m^2_{-}$ bisector, is given a negative index $-b$.
For initially produced $D^0$ ($\overline{D}^0$) mesons, if~the $C\!P$ conserving amplitudes and mixing parameters are small, the~expected ratios $R^+_{bj}$ ($R^-_{bj}$) of the number of decays in each negative Dalitz-plot bin ($-b$) to its positive counterpart ($+b$), for~each decay-time interval $j$, is
\begin{equation}
    R^\pm_{bj} \approx \frac{ r_b + r_b\frac{\left \langle t^2 \right \rangle_j}{4} \operatorname{Re}(z^2_{C\!P}-\Delta z^2) + \frac{\left \langle t^2 \right \rangle_j}{4}\left| z_{C\!P} \pm \Delta z \right|^2 + \sqrt{r_b}\left \langle t \right \rangle_j \operatorname{Re}[X_b^*(z_{C\!P} \pm \Delta z)] }{ 1 + \frac{\left \langle t^2 \right \rangle_j}{4} \operatorname{Re}(z^2_{C\!P}-\Delta z^2) + r_b\frac{\left \langle t^2 \right \rangle_j}{4}\left| z_{C\!P} \pm \Delta z \right|^2 + \sqrt{r_b}\left \langle t \right \rangle_j \operatorname{Re}[X_b(z_{C\!P} \pm \Delta z)] },
\label{bin_flip_Rbj}
\end{equation}
where $r_b$ is the value of $R^\pm_{bj}$ at decay time $t=0$, $X_b$ is the amplitude-weighted strong-phase difference between opposite bins and $\left\langle t \right\rangle_j$ ($\left\langle t^2 \right\rangle_j$) is the average (squared) decay time in each positive Dalitz-plot region in units of the $D^0$ lifetime.
The other parameters are defined according to
\begin{equation}
    z_{C\!P} \pm \Delta z \equiv -\left( \frac{q}{p} \right)^{\pm 1}(y+ix),
\label{bin_flip_z}
\end{equation}
and are obtained  from a simultaneous fit of the observed $R^\pm_{bj}$ ratios, where external information on $c_b \equiv \operatorname{Re}(X_b)$ and $s_b \equiv -\operatorname{Im}(X_b)$ is used as a constraint~\cite{Libby:2010nu,Ablikim:2020lpk}.
The $C\!P$-even mixing parameters $x_{C\!P}$ and $y_{C\!P}$ and the $C\!P$-violating parameters $\Delta x$ and $\Delta y$ are therefore~obtained
\begin{align}
    x_{C\!P} &\equiv -\operatorname{Im}(z_{C\!P}), \\
    y_{C\!P} &\equiv -\operatorname{Re}(z_{C\!P}), \\
    \Delta x &\equiv -\operatorname{Im}(\Delta z), \\
    \Delta y &\equiv -\operatorname{Re}(\Delta z).
\label{bin_flip_CPpars}
\end{align}

If $C\!P$ is conserved, $x_{C\!P} = x$, $y_{C\!P} = y$ and $\Delta x = \Delta y = 0$.

The data sample is split in two categories depending on whether the $K_s^0$ meson is reconstructed using long or downstream pions, as~already explained in Section~\ref{sec_KSKS}.
The flavour of the $D^0$ candidate is determined by the charge of the accompanying pion in the reconstructed $D^{*+} \to D^0 \pi^+$ decay.
To determine the signal yields, a~fit is performed on the invariant mass difference $\Delta m \equiv m(D^{*+}) - m(D^0)$ in each Dalitz-plot bin and decay-time interval, separately for $D^0$ flavour and $K_s^0$ category.
The fit uses an empirical model that takes into account contributions from signal and combinatorial background.
A total of about 31 million signal decays is found in the whole data~sample.

The signal selection includes requirements on the displacement and momenta of the $D^0$ decay products that introduce efficiency variations correlated between the phase-space coordinates and the $D^0$ decay time, resulting in a possible bias on the measurement.
A data-driven correction is therefore applied to make the decay-time acceptance uniform in the phase space, removing the correlation.
A further correction is performed to cancel detection asymmetries of the pions produced in the $D^0$ decays, since their kinematics depend on the Dalitz-plot coordinate and the $D^0$ flavour: the two-track $\pi^+ \pi^-$ detection asymmetry is evaluated by measuring the raw asymmetries in the control samples $D^+_s \to \pi^+ \pi^+ \pi^-$ and $D^+_s \to \phi \pi^+$ in each Dalitz-plot bin, after~weighting their kinematics to that of the signal~sample.

\begin{figure}[h]
\centering
\includegraphics[width=6.5 cm]{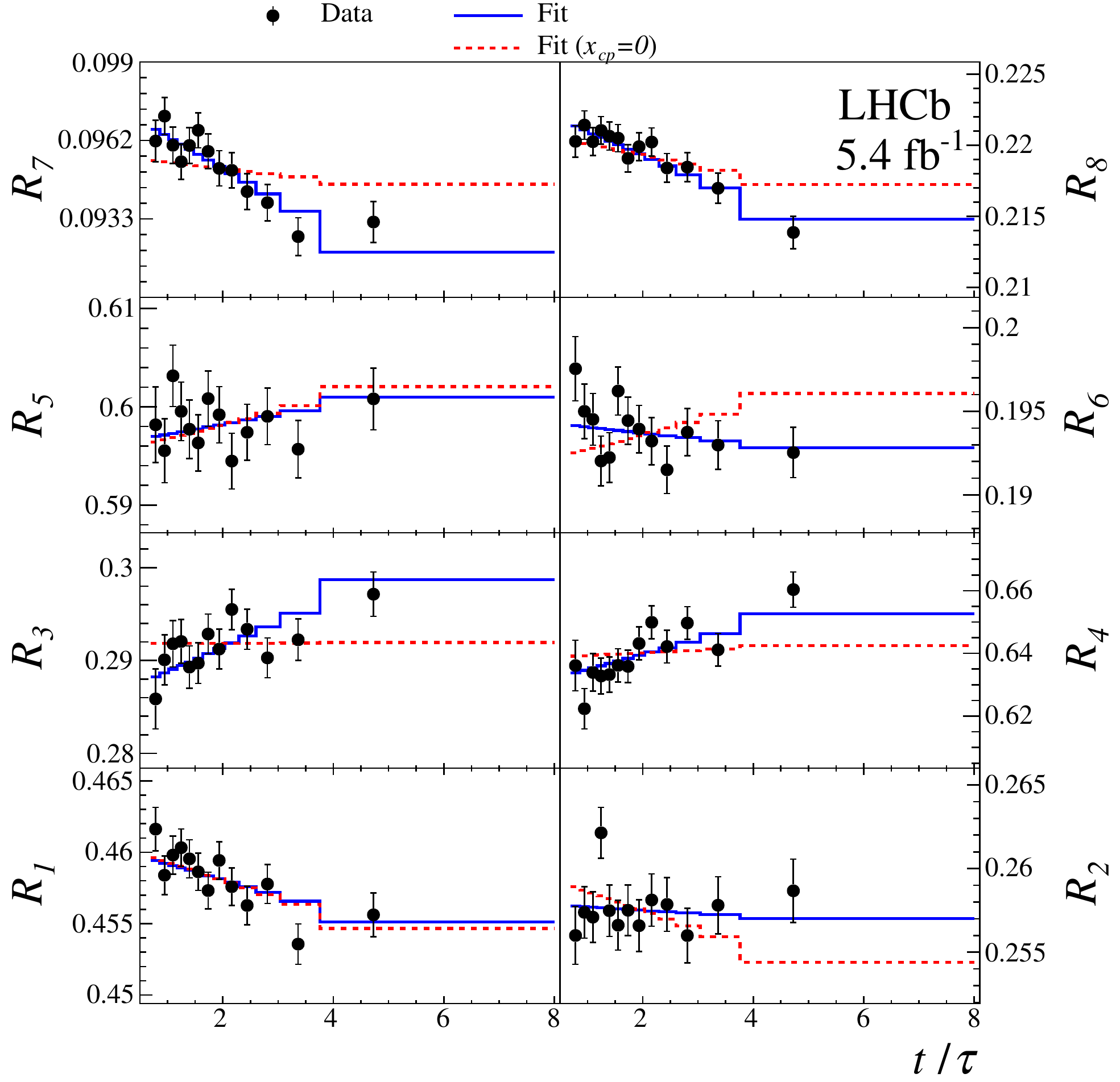}
\includegraphics[width=6.5 cm]{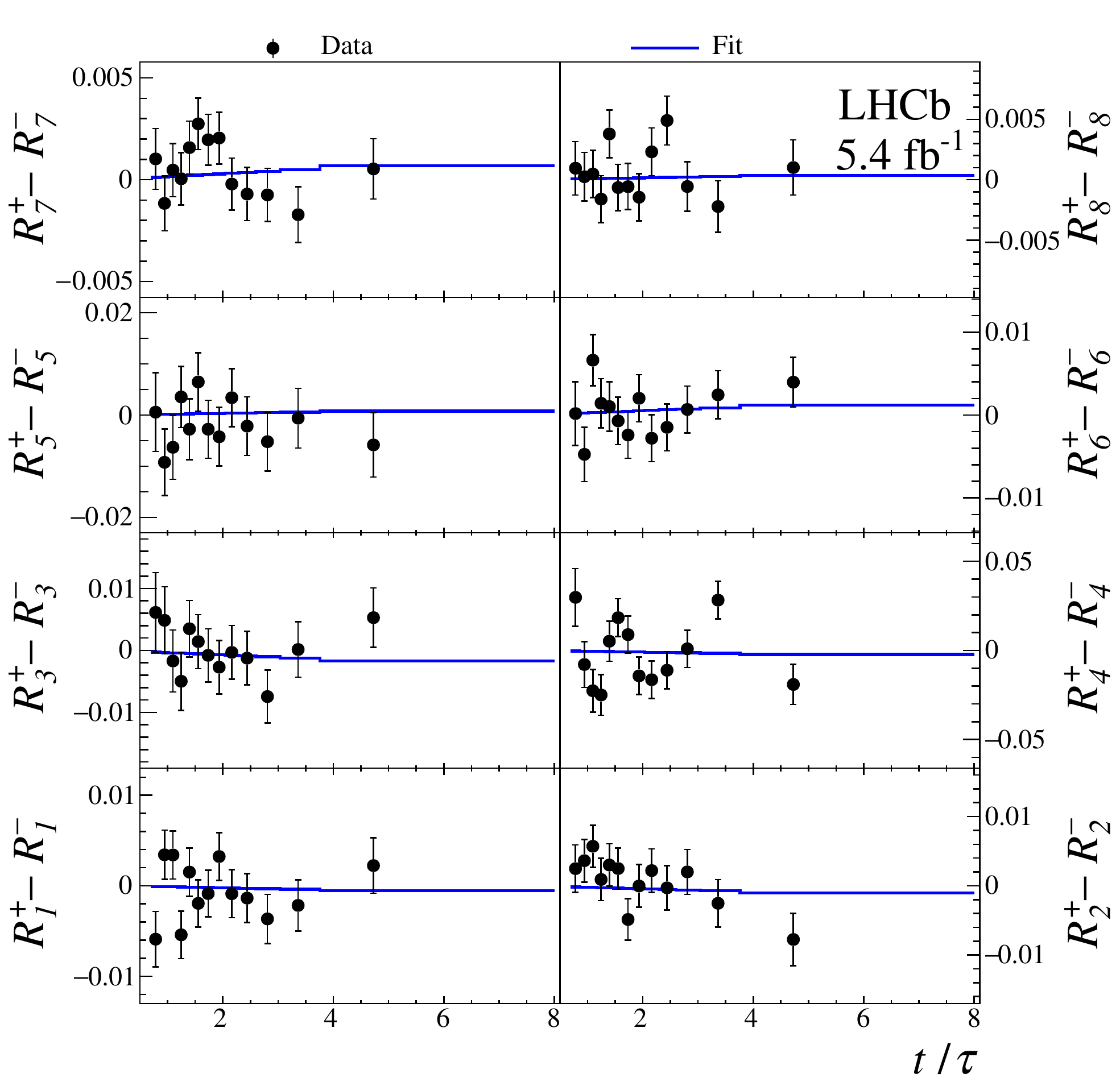}
\caption{(\textbf{Left}) $C\!P$-averaged yield ratios and (\textbf{right}) differences of $D^0$ and $\overline{D}^0$ yield ratios as a function of $D^0$ decay time, shown for each Dalitz-plot bin. Fit projections are~overlaid.}
\label{KSpipi_fits}
\end{figure}

A fit is performed on all the corrected $R^\pm_{bj}$ ratios to determine the mixing and $C\!P$ violating parameters.
The result is illustrated in Figure~\ref{KSpipi_fits}.
Various sources of systematic uncertainties are considered and assessed from ensembles of pseudoexperiments.
They take into account contributions related to reconstruction and selection effects, decay-time and $m_\pm$ resolution, presence of secondary $D^{*+}$ decays, $\pi^+\pi^-$ detection asymmetry, fit model and approximation of strong phase input to be constant in each bin.
The mixing and $C\!P$ violation parameters are measured to be
\begin{align*}
    x_{C\!P} &= ( 3.97 \pm 0.46 \pm 0.29 ) \times 10^{-3},\\
    y_{C\!P} &= ( 4.59 \pm 1.20 \pm 0.85 ) \times 10^{-3},\\
    \Delta x &= ( -0.27 \pm 0.18 \pm 0.01 ) \times 10^{-3},\\
    \Delta y &= ( 0.20 \pm 0.36 \pm 0.13 ) \times 10^{-3},
\end{align*}
where the first uncertainties are statistical and the second systematic~\cite{Aaij:2021aam}.

A likelihood function of $x$, $y$, $|q/p|$ and $\phi$ is built from the results using a likelihood-ratio ordering that assumes the observed correlations to be independent of the true parameter values~\cite{Aaij:2013zfa}, whose best fit point is
\begin{align*}
    x &= (3.98^{+0.56}_{-0.54}) \times 10^{-3}, \\
    y &= (4.6^{+1.5}_{-1.4}) \times 10^{-3}, \\
    |q/p| &= 0.996 \pm 0.052, \\
    \phi &= 0.056^{+0.047}_{-0.051}.
\end{align*}

This result corresponds to the first observation of a non-zero value of the mass difference $x$ of neutral charm meson mass eigenstates with a significance of more than seven standard deviations.
This measurement significantly improves limits on $C\!P$ violation in mixing in the charm~sector.

\section{Conclusions and future prospects}
\label{sec_conclusions}

The most recent results from LHCb concerning $C\!P$ violation in charm sector have been summarised here.
The analysis of data collected during the LHC Run~2 allowed the most precise measurement of time-dependent $C\!P$ asymmetry in $D^0$ decays to date to be obtained and a non-zero mass difference between neutral charm-meson eigenstates to be observed.
Furthermore, the~most precise measurement of $C\!P$ asymmetry in $D^0 \to K_s^0 K_s^0$ decays has been performed.
$C\!P$ asymmetries in two-body decays of charged charm mesons in final states with light neutral mesons have been determined with precisions of the order of $10^{-2}$.
Many key measurements are still to be performed with the full data sets collected by LHCb, such as the determination of $\mathcal{A}_{C\!P}(D^0 \to K^+ K^-)$, which is expected to be measured with an uncertainty of the order of $7 \times 10^{-4}$, and~of $y_{C\!P}^{h^+ h^-}$, defined as the deviation from unity of the average of the effective decay widths of $D^0 \to h^+ h^-$ and $\overline{D}^0 \to h^+ h^-$ decays.

Starting from 2022, the~Upgrade I of the LHCb experiment will increase the total integrated luminosity to $23~\mathrm{fb}^{-1}$ ($50~\mathrm{fb}^{-1}$) by the end of its third (fourth) data-taking period~\cite{LHCb:2012doh,LHCbCollaboration:2014vzo}.
An Upgrade II of the LHCb experiment was recently proposed and, if~approved, it will allow the total integrated luminosity to be increased to $300~\mathrm{fb}^{-1}$~\cite{Aaij:2244311}.
With such a large data set, the~statistical uncertainty of $\Delta Y$ will be reduced to a value below $2\times 10^{-5}$, and~that of $\mathcal{A}_{C\!P}(D^0 \to K^+ K^-)$ will become smaller than $10^{-4}$ ~\cite{LHCb:2018roe}.
The statistical uncertainty of the mixing parameters measured with the $D^0 \to K_s^0 \pi^+ \pi^-$ decay mode is expected to decrease by an order of magnitude.
For what concerns the $C\!P$ asymmetry of the $D^0 \to K_s^0 K_s^0$ decay, its projected statistical uncertainty is expected to decrease by a factor 4, reaching $\mathcal{O}(10^{-3})$.
For all the mentioned decay modes, there are no systematic uncertainties that are known to have irreducible contributions which exceed the ultimate statistical precision.
The LHCb Upgrade I and II will therefore have a strong potential of probing the Standard Model contribution and an impressive power to characterise new physics contributions to $C\!P$ violation in the charm~sector.

\bibliographystyle{ieeetr}
\bibliography{my_bib}

\end{document}